\documentclass[12pt,letterpaper,twoside]{amsart}
\usepackage{amsmath,amsthm,amssymb}
\usepackage{epsfig}

\newtheorem{theorem}{Theorem}
\newtheorem{corollary}{Corollary}
\newtheorem{lemma}{Lemma}
\newtheorem{proposition}{Proposition}

\theoremstyle{remark}
\newtheorem*{rem}{Remark}

\def\bmath#1{\mbox{\boldmath$#1$}}

\DeclareMathOperator{\ad}{ad}
\DeclareMathOperator{\Exp}{Exp}

\DeclareMathOperator*{\tr}{tr}
\DeclareMathOperator*{\qdet}{qdet}
\DeclareMathOperator*{\Det}{DET}
\DeclareMathOperator{\sgn}{sgn}
\DeclareMathOperator*{\diag}{diag}

\DeclareMathOperator*{\Van}{Van}
\newcommand{\G}{\Gamma}

\newcommand{\Haar}{\text{Haar}}

\newcommand{\SN}{S(N)}

\newcommand{\CK}{\mathcal{K}}
\newcommand{\CS}{\mathcal{S}}
\newcommand{\CI}{\mathcal{I}}

\newcommand{\SMN}{S(M,N)}

\newcommand{\frp}{\mathfrak{p}}

\newcommand{\frk}{\mathfrak{k}}
\newcommand{\frg}{\mathfrak{g}}

\newcommand{\fra}{\mathfrak{a}}

\newcommand{\bfC}{\mathbf{C}}
\newcommand{\bfR}{\mathbf{R}}

\newcommand{\bfx}{\mathbf{x}}
\newcommand{\bfxi}{\bmath{\xi}}

\newcommand{\pab}{P^{(a,b)}}
\newcommand{\pAB}{P^{(A,B)}}
\newcommand{\pABp}{P^{(A+1,B+1)}}
\newcommand{\Kab}{K^{(a,b)}}
\newcommand{\Iab}{I^{(a,b)}}
\newcommand{\Sab}{S^{(a,b)}}
\newcommand{\SabT}{S^{(a,b)T}}
\newcommand{\Dab}{D^{(a,b)}}
\newcommand{\KAB}{K^{(A,B)}}

\begin{document}

\title[Matrix ensembles associated to symmetric spaces]
{Random matrix ensembles\\ associated to compact symmetric spaces}

\email{eduenez@math.jhu.edu}

\author{Eduardo Due{\~n}ez}

\address{American Institute of Mathematics \and The Johns Hopkins
  University\\3400 N.\ Charles St.\\Baltimore, MD 21218}

\thanks{I wish to thank Prof.~Peter Sarnak for his continued
  encouragement and guidance as my Ph.~D. thesis advisor as well as
  Brian Conrey for making my stay at AIM possible.  This research has
  been supported in part by the FRG grant DMS--00--74028 from the NSF}

\begin{abstract}
  We introduce random matrix ensembles that correspond to the infinite
  families of irreducible Riemannian symmetric spaces of type~I.  In
  particular, we recover the Circular Orthogonal and Symplectic
  Ensembles of Dyson, and find other families of (unitary, orthogonal
  and symplectic) ensembles of Jacobi type.  We discuss the universal
  and weakly universal features of the global and local correlations
  of the levels in the bulk and at the ``hard'' edge of the spectrum
  (i.~e., at the ``central points'' $\pm1$ on the unit circle).
  Previously known results are extended, and we find new simple
  formulas for the Bessel Kernels that describe the local correlations
  at a hard edge.
\end{abstract}

\maketitle

\section{Introduction}\label{sec:intro}
Local correlations between eigenvalues of various ensembles of random
unitary, orthogonal or symplectic matrices, in the limit when their
size tends to infinity, are known to exhibit universal behavior in the
bulk of the spectrum.  Dyson's ``Threefold Way''~\cite{Dys4} predicts
that this behavior is to be expected universally in the bulk of the
spectrum, depending only on the symmetry type of the ensemble
(unitary, orthogonal or symplectic).  Unfortunately, for general
ensembles this conjecture remains open, though in the unitary case
(modeled after the Gaussian Unitary Ensemble) the universality of the
local correlations has been proven for some classes of
families~\cite{MR98k:47097,MR1711036,B-I,MR2000e:42010}.  In the
orthogonal and symplectic cases the extension of results known for
Gaussian ensembles is technically more complicated but some more
recent work deals with families of such ensembles~\cite{Stoj}.  Most
of the focus has been on non-compact (Gaussian and the like) matrix
ensembles.  In the present article we study families of compact
(circular) ensembles including, in particular, Dyson's circular
ensembles: the~COE, CUE and CSE~\cite{MR26:1111}.  First we fit
Dyson's ensembles into the framework of the theory of symmetric
spaces, and then we proceed to associate a matrix ensemble to every
family of irreducible compact symmetric space (all of these are known
by the work of Cartan~\cite{Ca27b,Ca27c}).  The most well-known of
these are the families of classical orthogonal, unitary and symplectic
groups of matrices, for which questions about universality have known
answers~\cite{K-S}.  These are the so-called compact symmetric spaces
of type~II.  Zirnbauer~\cite{MR97m:58012}, on the other hand, has
constructed the ``infinitesimal'' versions of the other (type~I)
ensembles, namely their tangent spaces at the identity element, which
is enough to derive their eigenvalue measures.  We, however, construct
the ``global'' ensembles associated to the infinite families of
compact symmetric spaces of type~I in a very explicit manner analogous
to Dyson's description of his circular ensembles.

\begin{table}[htb]
\begin{tabular}{|c|l|l|}
\hline
Type & G/K & Parameters \\
\hline\hline
\begin{tabular}{@{}c@{}}A~I\\(COE)\end{tabular}
& $U(R)/O(R)$ & $\beta=1$ (not Jacobi) \\
\hline
\begin{tabular}{@{}c@{}}A~II\\(CSE)\end{tabular}
& $U(2R)/USp(2R)$ & $\beta=4$ (not Jacobi)\\
\hline
A~III & $U(2R+L)/U(R+L)\times U(R)$ & $\beta=2, (a,b)=(L,0)$ \\
\hline
BD~I & $O(2R+L)/O(R+L)\times O(R)$ & $\beta=1, (a,b)=(\frac{L-1}2,-\frac12)$ \\
\hline
& $SO(4R)/U(2R)$ & $\beta=4, (a,b)=(0,0)$\\
\cline{2-3}
\raisebox{1.2ex}[0pt]{D~III} & $SO(4R+2)/U(2R+1)$ & $\beta=4, (a,b)=(2,0)$ \\
\hline
C~I & $USp(2R)/U(R)$ & $\beta=1, (a,b)=(0,0)$ \\
\hline
C~II & 
\begin{tabular}{@{}c@{}}$USp(4R+2L)/$\\$USp(2R+2L)\times USp(2R)$ \end{tabular}
& $\beta=4, (a,b)=(2L+1,1)$ \\
\hline
\end{tabular}
\caption{Parameters of the probability measure of the eigenvalues for ensembles of type~I.}
\label{tab:prob-eigen}
\end{table}

Besides Dyson's~COE and~CSE, the other compact matrix ensembles of
type~I are Jacobi ensembles in the sense that their joint eigenvalue
measure is given by
\begin{equation}
  \label{eq:90}
  d\nu(x_1,\dots,x_R)\propto
  \prod_{1\leq j<k\leq R} 
  |x_j-x_k|^\beta\prod_{j=1}^R(1-x_j)^a(1+x_j)^b dx_j\quad\text{on $[-1,1]^R$}
\end{equation}
for some parameters $a,b>-1$ (depending on the ensemble, see
table~\ref{tab:prob-eigen}) and $\beta=1,2,4$ (the ``symmetry
parameter'') in the orthogonal, unitary and symplectic cases,
respectively.  Here, the ``free'' eigenvalues are $x_j\pm \sqrt{-1}y_j$
---excluding eigenvalues equal to~$+1$ forced by the symmetry built
into the ensemble---.  Also, $R$ stands for the rank of the
corresponding symmetric space, and our interest is in the
semiclassical limit of the eigenvalue statistics as $R\to\infty$ ($L\geq0$ is
a fixed parameter: different values of $L$ yield different ensembles.)
The name ``Jacobi ensembles'' comes from the intimate connection
between the measure~\eqref{eq:90} and the classical Jacobi polynomials
on the interval $[-1,1]$.

Afterwards, we prove the universality of the local correlations for
general unitary, orthogonal and symplectic Jacobi ensembles (previous
results of Nagao and Forrester~\cite{N-F} are insufficient for our
purposes).  We rely on work of Adler \emph{et~al}~\cite{A-}.  At the
``hard edges'' $\pm1$ of the interval, Dyson's universality breaks down
and we obtain simple formulas for the Bessel kernel in terms of which
the hard edge correlations are expressed.  In a nutshell, for Jacobi
ensembles:
\begin{itemize}
\item Away from the ``hard edge'' $x=\pm1$, the local correlations
  follow the universal law of the GOE ($\beta=1$), GUE ($\beta=2$) or GSE
  ($\beta=4$).  Namely, in terms of local parameters $\xi_j$ around a
  fixed $z_o\in(-1,1)$ so that $x_j=\cos(\alpha_o+(\pi/R)\xi_j)$
  ($z_o=\cos\alpha_o$), these local correlations are given by
  \begin{equation}
    \label{eq:92}
    L_\beta^{(n)}(z_o;\xi_1,\dots,\xi_n) = \Det(\bar K_\beta(\xi_j,\xi_k))_{n\times n},
  \end{equation}
  where $\Det$ stands for either the usual ($\beta=2$) or quaternion
  ($\beta=1,4$) determinant, and $K_\beta$ is the (scalar or quaternion)
  Sine kernel (cf., equations~\eqref{eq:60}--\eqref{eq:62a}.)
\item At the hard edge $z_o=+1$, the local correlations depend on the
  parameter~$a$ of the Jacobi ensemble as well as on~$\beta$.  In terms
  of local parameters $\xi_j>0$ with $x_j=\cos((\pi/R)\xi_j)$ the same
  expression~\eqref{eq:92} holds except that the kernel $\bar K_\beta$ is
  to be replaced by a Bessel kernel $\hat K_\beta^{(a)}(\xi,\eta)$ given by
  equations~\eqref{eq:121}--\eqref{eq:63}.  At the hard edge $z_o=-1$
  the result is obtained by replacing~$a$ by~$b$.
\end{itemize}

\section{Dyson's Circular Ensembles as Symmetric Spaces}
\label{sec:circ-ens}
For motivational purposes we start by reviewing the construction of
the circular ensembles of Dyson and their probability measures of the
eigenvalues in a manner in which the theory of Riemannian symmetric
spaces is brought into play.

The~\emph{Circular Unitary Ensemble}~(CUE) is the set $S=\SN$ of all
$N\times N$ unitary matrices $H$, endowed with the unique probability
measure $d\mu(H)$ that is invariant under left (also right)
multiplication by any unitary matrix.  This requirement makes the
measure invariant under unitary changes of bases, hence the ensemble's
name.

In the study of statistics of eigenvalues, the relevant probability
measure is the one induced by $d\mu(H)$ on the torus $A=A(N)\subset\SN$
consisting of unitary diagonal matrices
\begin{equation}
  \label{eq:2}
  A=\{\diag(\lambda_1=e^{i\theta_1},\dots,\lambda_N=e^{i\theta_N})\},
\end{equation}
where $\Theta=(\theta_1,\dots,\theta_N)\in[0,2\pi)^N$, say.

To be more precise, let us denote by $K=K(N)$ the unitary group of $N\times
N$ matrices (its underlying set is just $\SN$).  Then we have a
surjective mapping
\begin{eqnarray}
  \label{eq:3}
  K\times A & \twoheadrightarrow & S \notag \\
  (k,a) & \mapsto & H=ka k^{-1},
\end{eqnarray}
and correspondingly there exists a probability measure $d\nu(a)$ on
$A$ such that, for any continuous function $f\in C(S)$,
\begin{equation}
  \label{eq:8}
  \int_S f(H) d\mu(H) = \int_K\int_Af(ka k^{-1})d\nu(a)d\Haar(k),
\end{equation}
where we denote by $d\Haar(k)$ the unique translation-invariant
probability measure on $K$ (so here $d\Haar=d\mu$).  This measure
$d\nu(a)$ can be pulled back to some measure on the space $[0,2\pi)^N$
of angles $\Theta$ which, abusing notation, we denote by $d\nu(\Theta)$.  The
measure $d\nu(\Lambda)$ (or $d\nu(\Theta)$) is the so-called probability measure
of the eigenvalues (for the CUE).  We have~\cite{Meh-rm},
\begin{equation}
  \label{eq:6}
  d\nu(\Theta) \propto \left|\Van(e^{i\Theta})\right|^2d\Theta \quad\text{on $[0,2\pi)^N$}.
\end{equation}
Here the symbol
``$\propto$'' stands for proportionality up to a constant (depending only
on $N$), $d\Theta = d\theta_1\dots d\theta_N$ is the usual translation-invariant
measure on the space of angles $\Theta$,
$e^{i\Theta}=(e^{i\theta_1},\dots,e^{i\theta_N})$ and, for a vector $\mathbf
x=(x_1,\dots,x_N)$, $\Van(\mathbf x)$ is the Vandermonde determinant
\begin{equation}
  \label{eq:10}
  \Van(\mathbf x) = \det_{N\times N}(x_j^{k-1}) = \prod_{1\leq j<k\leq N}(x_k-x_j).
\end{equation}

The construction of the \emph{Circular Orthogonal Ensemble}~(COE) is
as follows.  One starts with the set $S=\SN$ of $N\times N$ symmetric
unitary matrices $H$.  However, because $\SN$ is not a group, the
choice of the probability measure $d\mu(H)$ is not as obvious as it was
for the CUE.  Let $G=G(N)$ again be the group of $N\times N$ unitary
matrices $g$, and $K=K(N)\subset G(N)$ be the group of orthogonal matrices.
Let $\Omega(g)=(g^T)^{-1}$ be the involution of $G$ whose fixed-point set
is $K$.  Then we may identify
\begin{eqnarray}
  \label{eq:4}
  G/K &\simeq& S \notag \\
  G \ni g & \mapsto & H=g\Omega(g)^{-1}=:g^{1-\Omega},
\end{eqnarray}
and by general principles the translation-invariant probability
measures on $G$ and $K$ determine a unique $G$-invariant measure
$d\mu(\bar g)=d\mu(H)$ on $G/K\simeq S$ which satisfies
\begin{equation}
  \label{eq:24}
  \int_Gf(g)d\Haar(g) = \int_{G/K}\left(\int_K f(gk)d\Haar(k)\right)d\mu(\bar g),
\end{equation}
where on the right-hand side $g$ stands for a choice of an element
$g\in G$ such that $gK=\bar g$.  The left translation-invariance of
$d\Haar(g)$ ensures that $d\mu(\bar g)$ is invariant under left
translations by elements of $K$, therefore the measure $d\mu(H)$ is
invariant under orthogonal changes of bases, hence the ensemble's
name.

The probability measure of eigenvalues $d\nu(a)=d\nu(\Theta)$ is again that
which satisfies~\eqref{eq:8} (with the same torus $A\subset S$ as for the
CUE).  It is known that~\cite{MR26:1111}
\begin{equation}
  \label{eq:11}
  d\nu(\Theta) \propto |\Van(e^{i\Theta})|d\Theta \quad\text{on $[0,2\pi)^N$.}
\end{equation}

The constructions of the \emph{Circular Symplectic Ensemble}~CSE and
of its measure on eigenvalues $d\nu(\Theta)$ are very similar to the case
of the COE.  Here $S(N)$ consists of $2N\times2N$ self-dual unitary
matrices.  Namely, letting
\begin{equation}
  \label{eq:j-n}
  J=J_N=
  \begin{pmatrix}
     & -I_N \\
    I_N & 
  \end{pmatrix},
\end{equation}
then a matrix $H$ is self-dual if it equals its dual $H^D:=JH^TJ^T$.
If we let $G=G(N)$ be the group of $2N\times2N$ unitary matrices and
$K=K(N)$ be the subgroup of symplectic matrices $k$ (they satisfy
$kJk^T=J$) then $K$ is the fixed-point set of the involution
$\Omega(g)=(g^D)^{-1}$.  The identification~\eqref{eq:4} continues to hold
and~\eqref{eq:24} again defines the probability measure
$d\mu(H)=d\mu(\bar g)$ of the ensemble.  It is invariant under
symplectic changes of bases.

The torus $A$ consists here of diagonal matrices:
\begin{equation}
  \label{eq:14}
  A=\{\diag(e^{i\theta_1},\dots,e^{i\theta_N},e^{i\theta_1},\dots,e^{i\theta_N})\}
\end{equation}
with twice-repeated eigenvalues.  Then the probability measure of the
eigenvalues is characterized by~\eqref{eq:8}, and indeed
\begin{equation}
  \label{eq:15}
  d\nu(\Theta) \propto |\Van(e^{i\Theta})|^4d\Theta \quad\text{on $[0,2\pi)^N$.}
\end{equation}

Summing up, the measure on eigenvalues for the circular ensembles is
given by
\begin{equation}
  \label{eq:7}
  d\nu(\Theta) \propto |\Van(e^{i\Theta})|^\beta d\Theta,
\end{equation}
where $\beta=1,2,4$ in the orthogonal, unitary and symplectic cases,
respectively.

\begin{rem}
  It can be appreciated that the parameter $\beta$ determines the
  strength of the repulsion between nearby eigenvalues: this repulsion
  is stronger the larger $\beta$ is.  Hence anything that measures the
  local interactions between eigenvalues is likely to depend on $\beta$.
  This is the case, in particular, of the ``local correlations''
  between eigenvalues, cf.~section~\ref{sec:univ-local-corr}.
\end{rem}

\begin{rem}
  The apparent dissimilarity in the construction of the measure
  $d\mu(H)$ in the case of the unitary vs.~the orthogonal and
  symplectic ensembles is not essential.  In fact, the unitary
  ensemble $S(N)$ is still a quotient $G(N)/K(N)$ where $G(N)=U(N)\times
  U(N)$ is the direct product of two copies of the unitary group, and
  $K(N)$ is the diagonal of $G(N)$ (isomorphic to the unitary group
  itself).  If we identify $S(N)$ with the ``anti-diagonal''
  $\{H=(g,g^{-1})\}\subset G(N)$ and take $\Omega(g,h)=(h,g)$ then the
  construction of the ensemble and of the measures $d\mu(H)$ and
  $d\nu(\Theta)$ follows through in essentially the same manner.  We omit
  the details.  The key observation is that the constructions above
  show that the circular ensembles are examples of Riemannian globally
  symmetric spaces.
\end{rem}

\section{Compact Symmetric Spaces as Matrix Ensembles}
\label{sec:comp-symm-spac}
Any Riemannian globally symmetric space $X$ is locally isometric to a
product of irreducible ones (the symbol ``$\approx$'' means ``is locally
isometric to''):
\begin{equation}
  \label{eq:17}
  X\approx \prod_i X_i^{(c)}\times \prod_j X_j^{(nc)}\times E^\ell,
\end{equation}
where the $X_i^{(c)}$ (resp.,~the $X_j^{(nc)}$) are irreducible
symmetric spaces of compact (resp.,~non-compact) type, and
$E^\ell=(E^1)^\ell$ is $\ell$-dimensional Euclidean space (a flat manifold).
In the case of the circular ensembles, we have
\begin{eqnarray}
  \label{eq:16}
  \text{CUE} &=& U(N) \approx SU(N)\times S^1 \notag \\
  \text{COE} &=& U(N)/O(N) \approx (SU(N)/SO(N))\times S^1 \notag \\
  \text{CSE} &=& U(2N)/USp(2N) \approx (SU(2N)/USp(2N))\times S^1
\end{eqnarray}
where in each case the first factor is an irreducible symmetric space
of the compact type and the other (Euclidean) factor is a circle
$S^1\approx E^1$ (we write $S^1$ rather than $E^1$ to emphasize that the
spaces are compact).  In the language of differential geometry, the
probability measure of a circular ensemble is the one determined by
the natural volume element of the manifold.  Hence the natural
question arises as to how to construct a random matrix ensemble
corresponding to each (infinite) family of irreducible symmetric
spaces of compact type.  The restriction to infinite families is due
to the need to have a large parameter $N$ such that the number of
eigenvalues grows with $N$, and then we are interested mainly in
limiting statistics.

The presence of the Euclidean factor $S^1$ (which comes from the
subset of scalar multiples of the identity matrix within the ensemble)
is rather convenient and natural.  If we were to define
``irreducible'' circular ensembles analogously to Dyson's circular
ensembles, except requiring that they consist of matrices with unit
determinant, then the spaces so obtained would be irreducible
symmetric spaces of the compact type (i.~e., the factors $S^1$ would
disappear from~\eqref{eq:16}).  However, the measure on eigenvalues
would no longer be translationally invariant (under transformations of
the form $\Theta\mapsto\Theta+(t,\dots,t)$).  Namely, instead of the
measure~\eqref{eq:7}, we would obtain an asymmetric version given by
the same formula but with $\Theta$ replaced by
$\Theta=(\theta_1,\dots,\theta_{N-1},-\theta_1-\dots-\theta_{N-1})$ and with $d\theta_N$
omitted from the volume element $d\Theta$.  As may be expected from such a
loss of symmetry, a rigorous analysis of these ``irreducible''
ensembles would be more involved.

Since we are considering only compact symmetric spaces, it is possible
to normalize the natural volume element to obtain a probability
measure.  This is not the case for symmetric spaces of non-compact
type.  To clarify the difference, we analyze the example of the
classical Gaussian matrix ensembles, which also fit within the
framework of the theory of symmetric spaces (the construction is
analogous to that of the circular ensembles):
\begin{equation}
  \label{eq:18}
  \begin{split}
    &\text{GUE} \approx SL(N,\mathbf C)/SU(N) \times E^1 \\
    &\text{GOE} \approx SL(N,\mathbf R)/SO(N) \times E^1 \\
    &\text{GSE} \approx SU^*(2N)/USp(2N) \times E^1.    
  \end{split}
\end{equation}
Finding the probability measure on eigenvalues also reduces to a
factorization of measures $d\mu(H)=d\Haar(k)d\nu(a)$ in the sense
of~\eqref{eq:8}, where $K$ is still the group of invariance
(orthogonal, unitary, symplectic) of the ensemble's measure, but where
$A\simeq E^N$ is now a Euclidean space, which in the case of these
ensembles consists of real diagonal matrices which can be parametrized
by $N$-tuples $\Lambda=(\lambda_1,\dots,\lambda_N)$ of real numbers.  However, the
measure $d\mu(H)$ is certainly not the one obtained from the Riemannian
volume element $d\Haar(g)$ of $G$ through~\eqref{eq:11} since the
latter is not normalizable.  A choice has to be made to make this
measure into a finite one while preserving its left and right
$K$-invariance.  One possibility is provided by a ``Gaussian''
probability measure on $G$ proportional to
\begin{equation}
  \label{eq:22}
  e^{-\frac\beta2\tr g^2}d\Haar(g)
\end{equation}
(the symmetry parameter $\beta=1,2,4$ corresponds to the orthogonal,
unitary and symplectic cases, respectively, just as in the case of the
Orthogonal ensembles), which in turn yields the measure on
eigenvalues:
\begin{equation}
  \label{eq:20}
  d\nu(a) \propto e^{-\beta\sum\lambda_j^2}|\Van(\Lambda)|^\beta d\Lambda.
\end{equation}
It can be rightfully argued that the choice of the Gaussian
normalization for the measure on these matrix ensembles is rather
arbitrary and motivated by analytical rather than conceptual
considerations.  The point we wish to state here is that making such a
choice is unavoidable.  For the compact spaces, however, no such
choice needs to be made since their volume element already determines
a unique probability measure.  We will henceforth restrict our
attention to compact ensembles for that reason.

The general definition of a Riemannian symmetric space of the compact
type is as follows.  We start with a compact semisimple Lie algebra
$\frg$ (i.~e., $\exp(\ad(\frg))\subset GL(\frg)$ is compact) having an
involutive automorphism $\omega$.  Then $\frg$ splits into the sum of the
$(+1)$- and $(-1)$-eigenspaces of $\omega$ as
\begin{equation}
  \label{eq:21}
  \frg = \frk \oplus \frp.
\end{equation}
(the subspace $\frp\subset\frg$ can be identified with the tangent space to
$G/K$ at the identity coset $o=K/K$). $G/K$ is called a Riemannian
symmetric space of the compact type if
\begin{enumerate}
\item $K\subset G$ are Lie groups ($G$ connected).  Their Lie algebras are
  $\frk,\frg$; and
\item there is a (necessarily unique) involutive automorphism $\Omega$ of
  $G$ such that $(G^\Omega)_o \subset K \subset G^\Omega$, where $G^\Omega$ is the fixed-point
  set of $\Omega$ in $G$ (a Lie subgroup of $G$) and $(G^\Omega)_o$ is its
  identity component (then $d\Omega_e=\omega$).
\end{enumerate}

The complete list of irreducible symmetric spaces (up to local
isometry) is known by the classical work of Cartan.  As we will
explain later, it suffices to consider one matrix ensemble in each
equivalence class of locally isometric symmetric spaces, because the
measures on eigenvalues for locally isometric ensembles are the same.

The irreducible symmetric spaces of compact type are classified into
spaces of ``Type~I'' and ``Type~II''.  Of these the latter are
simplest to describe: they are the (connected) simple compact Lie
groups $G$, provided with a bi-invariant (under both left and right
translations) Riemannian metric.  Proving that such a $G$ is a
\emph{bona fide} symmetric space of the compact type as defined before
involves expressing it as $(G\times G)/G$ in a manner analogous to what we
did at the end of section~\ref{sec:circ-ens} for the CUE.

\begin{table}[htp]
\begin{tabular}{|l|l|c|}\hline
Type    &       $G/K$           &       Rank $R$     \\
\hline \hline
A~I     &       $SU(N)/SO(N)$   &       $N-1$     \\
A~II    &       $SU(2N)/USp(2N)$  &       $N-1$     \\
A~III   &       $SU(M+N)/S(U(M) \times U(N))$   &       $\min(M,N)$     \\
BD~I    &       $SO(M+N)/SO(M) \times SO(N)$    &       $\min(M,N)$     \\
D~III   &       $SO(2N)/U(N)$   &       $\lfloor N/2\rfloor$         \\
C~I     &       $USp(2N)/U(N)$    &       $N$     \\
C~II    &       $USp(2M+2N)/USp(2M) \times USp(2N)$    &       $\min(M,N)$  \\
\hline
\end{tabular}
\caption{The infinite families of symmetric spaces of type~I.}
\label{tbl:type-I}
\end{table}

Up to local isometry, the infinite families of Type~II spaces are
those of orthogonal $SO(N)$, unitary $SU(N)$ and (compact) symplectic
$USp(2N)$ groups.  The random matrix theory of these spaces is
well-known~\cite{K-S}.

The Type~I spaces, on the other hand, are those symmetric spaces $G/K$
of the compact type with $G$ simple.  The bi-invariant Riemannian
metric on $G$ determines that on the quotient $G/K$.
Table~\ref{tbl:type-I} lists the infinite families of Type~I spaces,
up to local isometry.

\emph{Without loss of generality, we assume henceforth that
  $\min(M,N)=N$.}

Choose a maximal abelian subalgebra $\fra$ of $\frg$ contained in
$\frp$.  Then the subgroup $A=\exp(\fra)$ is a torus that projects
onto a totally flat submanifold $AK/K\subset G/K$ (a flat torus).  This
totally flat manifold is maximal, and its dimension is the rank $R$ of
the symmetric space $G/K$.  Thus, $R=\dim(AK/K)=\dim A = \dim \fra$.

Guided by the exposition in the previous section, it is reasonable to
regard as ensembles the symmetric spaces $G/K$ of type~I endowed with
their normalized Riemannian volume elements $d\mu(\bar g)$, which
satisfy~\eqref{eq:24}.  However, the elements of these ensembles are
not matrices but rather cosets $\bar g=gK\in G/K$.

\begin{theorem}\label{thm:mat-real}
  The infinite families of type~I ensembles $G/K$ can be realized as
  matrix ensembles $S$.  Indeed, \eqref{eq:4} maps $G/K$ bijectively
  onto a submanifold $S\subset G$, and $G$ is a classical group of
  matrices, hence $S$ is a space of matrices.  Under this
  correspondence, $AK/K\subset G/K$ is mapped onto the torus $A$.  The
  action of $K$ on $G/K$ by left translation corresponds to the
  conjugation $H\mapsto kHk^{-1}$ on matrices $H\in S$, and any $H\in S$ is
  conjugate to some $a\in A$ under this action.  Moreover, two matrices
  in $A$ are conjugate under $K$ if and only if they have the same
  eigenvalues.
\end{theorem}
The proof of the theorem is a long exercise in elementary linear
algebra.  We shall omit most of the details, which can be found
in~\cite{Due-01}.  In what follows we describe the explicit matrix
ensembles $S$ which are the images of the imbedding~\eqref{eq:4}.

In each case, we choose the involution $\Omega$ of $G$ so that its
fixed-point set is exactly $K$.  The cases of A~I (COE) and A~II (CSE)
have been discussed already.  We introduce some notation (recall that
$J_N$ is defined by equation~\eqref{eq:j-n}):
\begin{eqnarray}
J'_N & = &
\left(
        \begin{array}{cc}
                 & I_N \\
                I_N &
        \end{array}
\right)_{2N\times2N}, \label{eq:jp-n}\\
J_{MN} & = &
\left(
        \begin{array}{cc}
                J_M &  \\
                 & J_N
        \end{array}
\right)_{(2M+2N)\times(2M+2N)}, \label{eq:j-mn}\\
J'_{MN} & = &
\left(
        \begin{array}{cc}
                J'_M &  \\
                 & J'_N
        \end{array}
\right)_{(2M+2N)\times(2M+2N)}, \label{eq:jp-mn}\\
I'_{MN} & = &
\left(
        \begin{array}{cc}
                I_M &  \\
                 & -I_N
        \end{array}
\right)_{(M+N)\times(M+N)}.\label{eq:ip-mn}
\end{eqnarray}
The canonical bilinear antisymmetric matrix $J_n$ in the definition of
the compact symplectic group $USp(2n)$ will be taken to
be~\eqref{eq:j-n} in the case of ensembles with one parameter $N$
($n=N$), and~\eqref{eq:j-mn} in the case of ensembles with two
parameters $M,N$ ($n=M+N$).

\emph{A~III.}
Take $M\geq N\geq 1$ and $G(M,N)=U(M+N)$.  Then $K(M,N)=U(M)\times U(N)$ is
the fixed-point set of the involution
\begin{equation}
  \label{eq:27}
   g \mapsto g^\Omega := I' g I',
\end{equation}
with $I' = I'_{MN}$ as in~\eqref{eq:ip-mn}.

The symmetric space $U(M+N) / U(M)\times U(N)= SU(M+N) / S(U(M)\times
U(N))$ is realized as the matrix ensemble
\begin{multline}
  \SMN := \{H = GI' \mbox{ such that $G\in U(M+N)$}\\
  \mbox{is Hermitian of signature $(M,N)$}\},\label{eq:28}
\end{multline}
under the identification~\eqref{eq:4}.  A choice of the abelian torus
$A$ is given by
\begin{equation}
  \label{eq:AIII-A}
  A=\left\{
  \begin{pmatrix}
    1_{M-N} & & \\
    & \Re\Lambda_N & -\Im\Lambda_N \\
    & \Im\Lambda_N & \Re\Lambda_N
  \end{pmatrix}\right\}
\end{equation}
where $\Lambda_N=\diag(\lambda_1,\dots,\lambda_N)$ is an arbitrary
diagonal unitary matrix.  Besides the eigenvalue $1$ with multiplicity
$M-N$, the eigenvalues of the matrix in~\eqref{eq:AIII-A} come in
$R=N$ pairs $\lambda_j,\lambda_j^{-1}$, $|\lambda_j|=1$.

\emph{BD~I.}  
Let $M\geq N\geq1$, $G(M,N)=O(M+N)$, and $K(M,N)=O(M)\times O(N)$
be the fixed-point set of the involution~\eqref{eq:27} with
$I'=I'_{MN}$ as in~\eqref{eq:ip-mn}.  Then $G/K=O(M+N)/O(M)\times
O(N)=SO(M+N)/S(O(M)\times O(N))\approx SO(M+N)/SO(M)\times SO(N)$ (the last two
spaces are locally isometric).
  
The symmetric space $O(M+N) / O(M) \times O(N)$ can be realized as the set
of matrices
\begin{multline}
  \SMN := \{H = gI' \mbox { such that $g \in O(M+N)$}\\
  \mbox{is symmetric of signature $(M,N)$} \},\label{eq:39}
\end{multline}
by means of~\eqref{eq:4}.  The torus $A$ is just as
in~\eqref{eq:AIII-A} and we get the same description for the
eigenvalues.

\emph{D~III.} Let $G(N) = SO(2N)$ and $K(N)=SO(2N)\cap Sp(2N,\mathbf C)\simeq
U(N)$:
\begin{equation}
  \label{eq:13}
  U(N)\ni g \mapsto 
  \begin{pmatrix}
    \Re g & -\Im g  \\
    \Im g & \Re g
  \end{pmatrix}\in K(N).
\end{equation}
Then $K(N)$ is the fixed-point set of the involution
\begin{equation}
  \label{eq:47}
  g \mapsto g^\Omega := J^T (g^{-1})^T J = J^T g J
\end{equation}
with $J=J_N$ as in~\eqref{eq:j-n}. We can identify $G(N)/K(N)$ with
the set
\begin{equation}
  \label{eq:48}
  \SN := \{H \in SO(2N)\mbox{ s.~t. $HJ$ is ``dexter'' antisymmetric} \}
\end{equation}
using equation~\eqref{eq:4}.  We now explain what we mean by a dexter
matrix.  Say $G$ is a $2N\times2N$ orthogonal antisymmetric matrix.  Then
an orthogonal change of basis puts it into the canonical form $J_N$.
However, this may not be possible by means of a \emph{proper}
orthogonal change of basis (i.~e., of determinant $+1$).
Specifically, when $N$ is even, the two complex structures $\pm J_N$ are
equivalent (under, say, the proper orthogonal change
of basis $J'_N$ as in~(\ref{eq:jp-n})), but when $N$ is odd they are not%
.  We call $G$ dexter if, by a proper orthogonal change of basis, it
can be taken into the canonical form $+J_N$.  Thus, for $N$ even, all
orthogonal antisymmetric matrices are dexter, whereas for $N$ odd,
only half of them are (in this case, conjugation by $J'_N$ takes
``dexter'' matrices into ``sinister'' ones and vice-versa).  Now, for
$H\in\SN$, $G := HJ$ is dexter antisymmetric, so our discussion above
proves the surjectivity of the mapping.

The torus $A$ is
\begin{equation}
  \label{eq:DIII-A}
  A = \left\{
  \begin{array}{l}
  \begin{pmatrix}
    \Re\Lambda_R&-\Im\Lambda_R&&\\
    \Im\Lambda_R&\Re\Lambda_R&&\\
    &&\Re\Lambda_R&\Im\Lambda_R\\
    &&-\Im\Lambda_R&\Re\Lambda_R\\
  \end{pmatrix}\qquad\mbox{for $N$ even;}\\
  \begin{pmatrix}
    1 & & & & & \\
    & \Re\Lambda_R&-\Im\Lambda_R&&\\
    &\Im\Lambda_R&\Re\Lambda_R&&\\
    &&&1&&\\
    &&&&\Re\Lambda_R&\Im\Lambda_R\\
    &&&&-\Im\Lambda_R&\Re\Lambda_R\\
  \end{pmatrix}\qquad\mbox{for $N$ odd.}
  \end{array}\right\},
\end{equation}
where $\Lambda_R=\diag(\lambda_1,\dots,\lambda_R)$ is a diagonal unitary matrix.
Besides the double eigenvalue $1$, which occurs for $N$ odd, the
matrices in~\eqref{eq:DIII-A} have $R$ quadruples of eigenvalues
$\lambda_j,\lambda_j,\lambda_j^{-1},\lambda_j^{-1}$.

\emph{C~I.} Here $G(N) = USp(2N)$, and $K(N) \simeq U(N)$ is the
fixed-point set of the involution~\eqref{eq:27} with $I' = I'_{NN}$ as
in~\eqref{eq:ip-mn}. Explicitly,
\begin{equation}
  \label{eq:26}
  U(N)\ni g \mapsto \left(
\begin{array}{cc}
        g & \\
        & (g^T)^{-1}
\end{array}
\right)\in K(N).
\end{equation}

Identify $G(N)/K(N)$ with the set
\begin{equation}
  \label{eq:56}
\SN := \{ H = GI' \mbox{ s.t. $G \in U(2N)$ is Hermitian and $JG =
  -\overline GJ$} \}
\end{equation}
by means of~\eqref{eq:4}.  The torus $A$ is
\begin{equation}
  \label{eq:CI-A}
  A = 
  \left\{ 
    \begin{pmatrix}
      \Re\Lambda_N&-\Im\Lambda_N\\
      \Im\Lambda_N&\Re\Lambda_N
  \end{pmatrix}
\right\},
  \end{equation}
with $\Lambda_N$ a unitary diagonal matrix as before.  The eigenvalues occur
in pairs just as in the case of~\eqref{eq:AIII-A} with $M=N$.

\emph{C~II.}  Let $M \geq N \geq 1$ and $G(M,N) = USp(2M+2N)$. We take the
complex structure $J = J_{MN}$ as in~\eqref{eq:j-mn}.  Then $K(M,N) =
USp(2M) \times USp(2N)$ consists exactly of those elements that also
stabilize
\begin{equation}
  \label{eq:64}
I' = \left(
\begin{array}{cc}
I'_{MN} & \\
 & I'_{MN}
\end{array}
\right),
\end{equation}
with $I'_{MN}$ as in~(\ref{eq:ip-mn}), so that $K(M,N)$ is the
fixed-point set of the involution
\begin{equation}
  \label{eq:65}
g \mapsto g^\Omega := I' g I'.
\end{equation}
We can realize the symmetric space $G(M,N)/K(M,N)$ as the set of
matrices
\begin{multline}
  \label{eq:66}
  \SMN := \{ H = G I' \mbox{ such that $G\in USp(2M+2N)$}\\
  \mbox{is Hermitian of signature $(M,N)$}\},
\end{multline}
where we mean the \emph{quaternionic} signature as discussed below.
We recall that any matrix $G\in Sp(2n,\bfC)$ which is Hermitian
($\overline{G} = G^T$), has real eigenvalues and can be diagonalized
with a symplectic matrix $g\in Sp(2n,\bfC)$, that is,
\begin{equation}
  \label{eq:67}
g^{-1}Gg = \left(
\begin{array}{cc}
\Delta_n & \\
 & \Delta_n^{-1}
\end{array}
\right)
\end{equation}
for some real diagonal matrix $\Delta_n$. The usual signature of $G$ is of
the form $(2a,2b)$, so we call $(a,b)$ the quaternionic signature.
The identification is, of course, given by~\eqref{eq:4}.  The torus
$A$ is
\begin{equation}
  \label{eq:CII-A}
  A = \left\{
  \begin{pmatrix}
    I_{M-N} & & & & & \\
    & \Re\Lambda_N & -\Im\Lambda_N & & & \\
    & \Im\Lambda_N & \Re\Lambda_N & & & \\
    & & & I_{M-N} & & \\
    & & & & \Re\Lambda_N & \Im\Lambda_N \\
    & & & & -\Im\Lambda_N & \Re\Lambda_N
  \end{pmatrix}\right\}
\end{equation}
with $\Lambda_N$ unitary diagonal.  Besides the eigenvalue $1$ with
multiplicity $2(M-N)$, the other eigenvalues occur in quadruples like
those of the matrices in~\eqref{eq:DIII-A}.

For each of the ensembles, the torus $A$, which has dimension equal to
the rank $R$ of the symmetric space, is parametrized by diagonal
unitary matrices
\begin{equation}
  \label{eq:29}
  \Lambda_R=\diag(\lambda_1,\dots,\lambda_R),\qquad |\lambda_j|=1.
\end{equation}
Abusing notation, we will also write $\Lambda_R$ for the vector
$(\lambda_1,\dots,\lambda_R)$.  The tangent space $\fra$ to this torus at the
identity is identified with the space of $R$-tuples
$i\Theta=(i\theta_1,\dots,i\theta_R)$, $\theta_j\in\bfR$.  Recall that we identify
$\frp$ with the tangent space to $G/K$ at the base-point $o=K/K$.  The
exponential maps $\Exp$ of $G/K$ and $\exp$ of $G$ are related by
\begin{equation}
  \label{eq:30}
  \Exp(X) = \exp(X)K \in G/K
\end{equation}
for $X\in\frp=T_o(G/K)$.  For $i\Theta\in\fra$, $\exp(i\Theta)$ is given by the
matrix on the right-hand side of equations~\eqref{eq:AIII-A},
\eqref{eq:DIII-A}, \eqref{eq:CI-A} and~\eqref{eq:CII-A}, respectively,
provided we choose $\lambda_j=e^{i\theta_j}$ in~\eqref{eq:29}.

\begin{proposition}[$KAK$ decomposition]\label{thm:KAK}
  Let $G/K$ be a symmetric space of the compact type and $A\subset G$ be as
  above.  The mapping
  \begin{eqnarray}
    \label{eq:23}
    K\times A\times K &\twoheadrightarrow& G \notag\\
    (k_1,a,k_2) &\mapsto& k_1ak_2
  \end{eqnarray}
  is a surjection.
\end{proposition}
The $KAK$ decomposition has an integral counterpart.
\begin{proposition}[Weyl's integration formula]\label{thm:Weyl-fmla}
  There is a measure $d\bar\nu(a)$ on $A$ such that, for any $f\in
  C(G)$,
  \begin{equation}
    \label{eq:32}
    \int_G f(g)dg = \int_K\int_K\int_A f(k_1ak_2) d\bar\nu(a) dk_2\,dk_1.
  \end{equation}
  (We have simplified our notation by dropping the name ``Haar'' of
  the respective invariant measures.)  Denote by $\Xi^+$ the set of
  positive roots of the symmetric Lie algebra $(\frg,\omega)$, and by
  $m_\alpha$ the multiplicity of a positive root $\alpha\in\Xi^+$.  Then
  \begin{equation}
    \label{eq:33}
    d\bar\nu(a) \propto \prod_{\alpha\in\Xi^+} |\sin\alpha(\Theta)|^{m_\alpha}da = \Delta(\Theta)da,
  \end{equation}
  say, where $\Theta$ is chosen so $a=\exp(i\Theta)$.
\end{proposition}
(With the notation above, we write $i\Theta=\log(a)$.  This $\Theta$ is
well-defined modulo $2\pi$.)

Now recall that the (positive) roots of $(\frg,\omega)$ are certain
non-zero real-valued linear functionals on $\fra$ (in fact one should
speak about the roots which are positive with respect to a fixed Weyl
chamber in $\fra$).  The root systems of the irreducible orthogonal
Lie algebras of compact type are well-known by Cartan's work.

\begin{proposition}\label{thm:mltpl}
  The positive roots and multiplicities for the irreducible orthogonal
  Lie algebras of type~I are as follows (let $L=M-N$ in the case of
  ensembles with two parameters).
  \begin{itemize}
  \item A~I.
    \begin{center}
      \begin{tabular}{|c|c|}
        \hline
        $\alpha$ & $m_\alpha$ \\
        \hline
        $\theta_k-\theta_j,\quad 1\leq j<k\leq R$ & $1$ \\
        \hline
      \end{tabular}
    \end{center}
  \item A~II.
    \begin{center}
      \begin{tabular}{|c|c|}
        \hline
        $\alpha$ & $m_\alpha$ \\
        \hline
        $\theta_k-\theta_j,\quad 1\leq j<k\leq R$ & $4$ \\
        \hline
      \end{tabular}
    \end{center}
  \item A~III.
    \begin{center}
      \begin{tabular}{|c|c|}
        \hline
        $\alpha$ & $m_\alpha$ \\
        \hline
        $\theta_k-\theta_j,\quad 1\leq j<k\leq R$ & $4$ \\
        \hline
      \end{tabular}
    \end{center}
  \item BD~I.
    \begin{center}
      \begin{tabular}{|c|c|}
        \hline
        $\alpha$ & $m_\alpha$ \\
        \hline
        $\theta_k\pm\theta_j,\quad 1\leq j<k\leq R$ & $1$ \\
        \hline
        $\theta_j,\quad 1\leq j\leq R$ & $L$ \\
        \hline
      \end{tabular}
    \end{center}
  \item D~III. $N$ even.
    \begin{center}
      \begin{tabular}{|c|c|}
        \hline
        $\alpha$ & $m_\alpha$ \\
        \hline
        $\theta_k\pm\theta_j,\quad 1\leq j<k\leq R$ & $4$ \\
        \hline
        $2\theta_j,\quad 1\leq j\leq R$ & $1$ \\
        \hline
      \end{tabular}
    \end{center}
  \item D~III. $N$ odd.
    \begin{center}
      \begin{tabular}{|c|c|}
        \hline
        $\alpha$ & $m_\alpha$ \\
        \hline
        $\theta_k\pm\theta_j,\quad 1\leq j<k\leq R$ & $4$ \\
        \hline
        $\theta_j,\quad 1\leq j\leq R$ & $4$ \\
        \hline
        $2\theta_j,\quad 1\leq j\leq R$ & $1$ \\
        \hline
      \end{tabular}
    \end{center}
  \item C~I.
    \begin{center}
      \begin{tabular}{|c|c|}
        \hline
        $\alpha$ & $m_\alpha$ \\
        \hline
        $\theta_k\pm\theta_j,\quad 1\leq j<k\leq R$ & $1$ \\
        \hline
        $2\theta_j,\quad 1\leq j\leq R$ & $1$ \\
        \hline
      \end{tabular}
    \end{center}
  \item C~II.
    \begin{center}
      \begin{tabular}{|c|c|}
        \hline
        $\alpha$ & $m_\alpha$ \\
        \hline
        $\theta_k\pm\theta_j,\quad 1\leq j<k\leq R$ & $4$ \\
        \hline
        $\theta_j,\quad 1\leq j\leq R$ & $4L$ \\
        \hline
        $2\theta_j,\quad 1\leq j\leq R$ & $3$ \\
        \hline
      \end{tabular}
    \end{center}
  \end{itemize}
\end{proposition}

We are now ready to derive the measure on eigenvalues for ensembles of
type~I.

\begin{theorem}
  The measure on eigenvalues for a symmetric space of type~I is given
  by
  \begin{equation}
    \label{eq:37}
    d\nu(a) \propto \Delta(\Theta/2)da
    = \prod_{\alpha\in\Xi^+} \left|\sin\frac12 \alpha(\Theta)\right|^{m_\alpha}da, \quad
    i\Theta=\log a.
  \end{equation}
\end{theorem}
Using Weyl's integration formula, we deduce that, for any $f\in C(S)$,
\begin{equation}
  \label{eq:34}
  \begin{split}
    \int_S f(H)d\mu(H) &= \int_{G/K}f((gk)^{1-\Omega})dk\,d\mu(\bar g) 
    \qquad\text{(by~\eqref{eq:4})}\\
    &= \int_G f(g^{1-\Omega})dg \quad\text{(since $k^{1-\Omega}=e$)}\\
    &= \int_K\int_A\int_K f((k_1ak_2)^{1-\Omega})dk_2d\bar\nu(a)dk_1 \\
    &= \int_K\int_A\int_K f((k_1a)^{1-\Omega})dk_2d\bar\nu(a)dk_1 \\
    &= \int_K\int_A f((ka)^{1-\Omega})d\bar\nu(a)dk \\
    &= \int_K\int_Af(ka^2k^{-1})d\bar\nu(a)dk. \qquad\text{(since $a^{1-\Omega}=a^2$)}
  \end{split}
\end{equation}
This ought to be compared with~\eqref{eq:8}, which defines the measure
$d\nu(\Lambda)$ on eigenvalues.  A key property of the measure $d\bar\nu(a)$
defined by~\eqref{eq:33} is reflected in the fact that
$\Delta(\Theta)=\Delta(\Theta')$ if $\Theta\equiv\Theta'\mod \pi$ (this follows in general from the
fact that the roots take integral values on the ``unit lattice''
$\exp^{-1}(e)$, and can be verified for ensembles of type~I directly
using proposition~\ref{thm:mltpl}).  From that observation, it follows
that:
\begin{equation}
  \label{eq:38}
  \begin{split}
    \int_K\int_Af(ka^2k^{-1})d\bar\nu(a)dk &\propto \int_K\int_{[0,2\pi]^R} f(k\exp(2i\Theta)k^{-1})\Delta(\Theta)d\Theta\,dk \\
    &= 2^R\int_K\int_{[0,\pi]^R}f(k\exp(2i\Theta)k^{-1})\Delta(\Theta)d\Theta\,dk\\
    &= \int_K\int_{[0,2\pi]^R}f(k\exp(i\Theta)k^{-1})\Delta(\Theta/2)d\Theta\,dk\\
    &= \int_K\int_Af(ka k^{-1})\Delta(\Theta/2)da\,dk.
  \end{split}
\end{equation}
When put together with~\eqref{eq:34}, this proves~\eqref{eq:37}.

Now we restrict attention to the most interesting case, that of
``class functions'' $f\in C(K\backslash S)$, that is, those functions on $S$
which depend only on the eigenvalues of the matrix, viz
\begin{equation}
  \label{eq:35}
  f(kak^{-1})=f(a).
\end{equation}

The tori~$A$ are parametrized by $R$-tuples $(\lambda_j=e^{i\theta_j})$.  From
the knowledge of the structure of the set of eigenvalues of the
matrices in these tori, we see that for all the ensembles of type~I
\emph{except} for Dyson's~A~I and~A~II, changing the sign of any
$\theta_j$ does not change the set of eigenvalues since these always come
in pairs $\{e^{\pm i\theta_j}\}$ (with single or double multiplicity), hence
any class function $f\in C(K\backslash S)$ is determined by its values on
$\exp([0,\pi]^R)\subset A$, and correspondingly
\begin{equation}
  \label{eq:36}
  \begin{split}
    \int_S f(H) d\mu(H) &= \int_Af(a)d\nu(a) \propto
    \int_{[-\pi,\pi]^R} f(\exp(i\Theta))\Delta(\Theta/2)d\Theta\\
    &= 2^R\int_{[0,\pi]^R} f(\exp(i\Theta))\Delta(\Theta/2)d\Theta.
  \end{split}
\end{equation}
Hence, except in the cases of~A~I and~A~II, it is convenient to regard
the measure on eigenvalues as one supported on $[0,\pi]^R$.  Noting
that the contribution of a pair of roots $\theta_k\pm\theta_j$ to $\Delta(\Theta/2)$ is
\begin{equation}
  \label{eq:43}
  \left|\sin\left(\frac{\theta_k-\theta_j}2\right)\sin\left(\frac{\theta_k+\theta_j}2\right)\right| 
  \propto |\cos\theta_k-\cos\theta_j|,
\end{equation}
it is clear that for all the ensembles of type~I, except for the~COE
and the~CSE, the measure on eigenvalues is proportional to the measure
\begin{equation}
  \label{eq:44}
  \prod_{1\leq j<k\leq R}|\Van(\cos\Theta)|^\beta \prod_{1\leq
    j\leq R}|\sin\theta_j|^P|\sin(\theta_j/2)|^Q d\Theta,\qquad\text{on $[0,\pi]^R$.}
\end{equation}
(Here $\beta=1,2,4$ according to the multiplicity $m_\alpha$ of the roots
$\theta_k\pm\theta_j$.)  Because $|\sin\theta|=|1-\cos\theta|^{1/2}|1+\cos\theta|^{1/2}$ and
$|\sin(\theta/2)|=2^{-1/2}|1-\cos\theta|^{1/2}$, the above is proportional to
the measure
\begin{equation}
  \label{eq:45}
    \prod_{1\leq j<k\leq R}|\Van(\cos\Theta)|^\beta \prod_{1\leq j\leq
    R}|1-\cos\theta_j|^p|1+\cos\theta_j|^q d\Theta,\qquad\text{on$[0,\pi]^R$.}
\end{equation}
We make the change variables $\Theta\mapsto \bfx=\cos\Theta$ to obtain
\begin{equation}
  \label{eq:46}
  d\nu(\bfx) \propto \prod_{1\leq j<k\leq R}|\Van(\bfx)|^\beta \prod_{1\leq j\leq R} 
  |1-x_j|^a|1+x_j|^b d\bfx,\qquad\text{on$[-1,1]^R$,}
\end{equation}
where $a=p-1/2$, $b=q-1/2$, and $d\bfx=dx_1\dots dx_R$.  The weight
function 
\begin{equation}
  \label{eq:49}
  w(x)=|1-x|^a|1+x|^b\qquad\text{on $[-1,1]$}
\end{equation}
is that with respect to which the classical Jacobi orthogonal
polynomials $\pab_n(x)$ are defined, so a matrix ensemble for which
the probability measure of the eigenvalues is given by~\eqref{eq:46}
is called a Jacobi ensemble (with parameters $(a,b)$).  For $\beta=1,2,4$
we call such an ensemble orthogonal, unitary or symplectic,
respectively.

Recall that, for the~COE and~CSE, the
probability measure of the eigenvalues is given by~\eqref{eq:7}.  It
coincides with that given by Weyl's formula
(proposition~\ref{thm:Weyl-fmla}) since
\begin{equation}
  \label{eq:50}
  |e^{i\theta_k}-e^{i\theta_j}| = 2\left|\sin\left(\frac{\theta_k-\theta_j}2\right)\right|.
\end{equation}

For completeness, table~\ref{tab:prob-eigen-II} is the analogue of
table~\ref{tab:prob-eigen} for (the infinite families of) symmetric
spaces of type~II (compact Lie groups).  The~CUE is a circular
ensemble with $\beta=2$ and measure on eigenvalues~\eqref{eq:7}, whereas
the orthogonal and symplectic groups are unitary Jacobi ensembles.
\begin{center}
\begin{table}[h]
\begin{tabular}{|l|l|l|l|}
\hline
Type & $\SN$ & Parameters \\
\hline\hline
$\mathfrak{a}_N$ (CUE)
& $U(N)$ & $\beta=2$ \\
\hline
$\mathfrak{b}_N$ & $SO(2N+1)$ & $\beta=2,(a,b)=(\frac12,-\frac12)$\\
\hline
$\mathfrak{c}_N$ & $USp(2N)$ & $\beta=2,(a,b)=(\frac12,\frac12)$ \\
\hline
$\mathfrak{d}_N$ & $SO(2N)$ & $\beta=2,(a,b)=(-\frac12,-\frac12)$ \\
\hline
\end{tabular}
\caption{Parameters of the probability measure of the eigenvalues for ensembles of type~II.}
\label{tab:prob-eigen-II}
\end{table}
\end{center}

\section{Universality of Local Correlations}
\label{sec:univ-local-corr}
In this section we analyze the limiting correlation functions for
general Jacobi ensembles.  As we have shown, with the exception of
Dyson's COE (A~I) and CSE (A~II), the ensembles of type~I are special
cases of (orthogonal, unitary or symplectic) Jacobi ensembles.

We consider the joint probability measure of the $R$ levels (we speak
about levels rather than eigenvalues since the natural variables to
use are $x_j=\Re\lambda_j$) given in the general form
\begin{equation}
  \label{eq:113}
  d\nu(\bfx_R) = P_R(\bfx_R)\,d\bfx_R,
\end{equation}
where $\bfx_R=(x_1,\dots,x_R)$ is an $R$-tuple of levels.  The
$n$-level correlation function $I_R^{(n)}(\bfx_n)$ is
defined by
\begin{equation}
\label{eq:87}
I_R^{(n)}(\bfx_n)=\frac{R!}{(R-n)!}\int\!\! \cdots \!\!\int
P_R(\bfx_n,x_{n+1},\dots,x_R)dx_{n+1}\cdots dx_R.
\end{equation}
It is, loosely speaking, the probability that $n$ of the levels,
regardless of order, lie in infinitesimal neighborhoods of $x_1,\ldots,x_n$
(but the total mass of the measure $I_R^{(n)}(\bfx_n)d\bfx_n$ is now
$R!/(R-n)!$ and not $1$).

The semi-classical limit $R\to\infty$ is of great interest.  The so-called
``universality conjecture'' (which dates back to the work of
Dyson~\cite{Dys4}) states that the \emph{local} correlations of the
eigenvalues in the bulk of the spectrum tend to very specific limits
that depend only on the symmetry parameter $\beta$.  Special cases of the
truth of this assertion are known.  In particular, in the unitary case
$\beta=2$, the result is proven in certain
generality~\cite{MR1702716,MR1711036,MR2000e:42010,B-I}, but for
$\beta=1,4$ it is known only for special ensembles such as the circular
ensembles of Dyson~\cite{MR26:1111,MR26:1112,MR26:1113} and, by work
of Nagao and Forrester~\cite{N-F}, for most Laguerre ensembles and
Jacobi ensembles.  However, the latter assumes that the parameters
$a,b$ are strictly positive, hence it is not applicable to ensembles
of type~I (cf., table~\ref{tab:prob-eigen}).

It is an extremely important fact that for general orthogonal, unitary
and symplectic ensembles the correlation functions can be expressed as
determinants (which discovery goes back, in the unitary case, to the
work of Gaudin and Mehta~\cite{Ga60,MR22:3741}, and in the orthogonal
and symplectic cases to Dyson's study of his circular ensembles, and
later extended by Chadha, Mahoux and
Mehta~\cite{MR83c:81032,MR83h:81026,MR92i:58027} to the general case).
In the case of unitary Jacobi ensembles there exists a scalar-valued
kernel $\Kab_{R2}(x,y)$ defined in terms of the classical Jacobi
orthogonal polynomials $\pab_n(x)$ (the projector kernel onto the span
of the first $R$ Jacobi polynomials) satisfying~\cite{N-W1}
\begin{equation}
  \label{eq:51}
  I_{R\beta}^{(n)}(\bfx_n) =\det(K_{R\beta}(x_j,x_k))_{j,k=1,\ldots,n}.
\end{equation}
In the case of the orthogonal (resp.,~symplectic) Jacobi ensembles,
there exists a matrix-valued kernel~\cite{N-W1} (alternatively, a
``quaternion'' kernel)
\begin{equation}
  \label{eq:52}
  \Kab_{R\beta}(x,y) =
  \begin{pmatrix}
    \Sab_{R\beta}(x,y) & \Iab_{R\beta}(x,y)-\delta\epsilon(x-y) \\
    \Dab_{R\beta}(x,y) & \SabT_{R\beta}(x,y)
  \end{pmatrix},
\end{equation}
where $\delta=1$ (resp., $\delta=0$---the $\epsilon$-term is absent in the
symplectic case),
\begin{equation}
  \label{eq:53}
  \epsilon(z)=\frac12\sgn(z)=\frac12\frac z{|z|},
\end{equation}
and the scalar kernel $\Sab_{R\beta}$ is defined in terms of the
skew-orthogonal polynomials of the second (resp.,~first) kind
depending on the weight~\eqref{eq:49} and the other quantities are
given by
\begin{align}
  \label{eq:85}
  \Iab_{R\beta}(x,y) &= -\int_x^y \Sab_\beta(x,z)dz, \\
  \label{eq:86}
  \Dab_{R\beta}(x,y) &= \partial_x \Sab_{R\beta}(x,y),  \\
  \label{eq:88}
  \SabT_{R\beta}(x,y) &= \Sab_{R\beta}(y,x).  
\end{align}
The matrix kernel~\eqref{eq:52} is self-dual in the sense that
$\Kab_{R\beta}(y,x) = \Kab_{R\beta}(x,y)^D$ (cf.,
section~\ref{sec:circ-ens}). The correlation functions themselves are
given by
\begin{equation}
  \label{eq:54}
  I_{R\beta}^{(n)}(\bfx_n) =\sqrt{\det(K_{R\beta}(x_j,x_k))_{n\times n}}.
\end{equation}
Indeed, if the matrix $(K_{R\beta}(x_j,x_k))_{n\times n}$ is interpreted as a
quaternion self-dual matrix~\cite{Meh-mt}, then the right-hand side
of~\eqref{eq:54} is its Dyson's ``quaternion determinant''
$\qdet$~\cite{MR26:1111,MR26:1112,MR26:1113}, so~\eqref{eq:54} can be
rewritten:
\begin{equation}
  \label{eq:55}
    I_{R\beta}^{(n)}(\bfx_n) =\qdet(K_{R\beta}(x_j,x_k))_{n\times n}.
\end{equation}
\begin{rem}
  In what follows we will sometimes unify notation by writing $\Det$
  (all caps) to signify the usual determinant when $\beta=2$ and the
  quaternion determinant when $\beta=1,4$.  Thus, equations~\eqref{eq:51}
  and~\eqref{eq:55} will be written
  \begin{equation}
    \label{eq:89}
    I_{R\beta}^{(n)}(\bfx_n) =\Det(K_{R\beta}(x_j,x_k))_{n\times n}. 
  \end{equation}
\end{rem}

The first quantity of interest is the (global) \emph{level density.}
Indeed, since the first correlation function has total mass $R$, one
might expect that the probability measure $R^{-1}I_R^{(1)}(x)dx$ on
$[-1,1]$ tend to a limiting measure as $R\to\infty$.  We define the level
density to be the corresponding probability density function:
\begin{equation}
  \label{eq:41}
  \rho(x) = \lim_{R\to\infty}R^{-1}I_R^{(1)}(x).
\end{equation}
Assuming $\rho(x)$ to be continuous, the bulk of the spectrum is the set
$\{x:\rho(x)>0\}$: points where the level density vanishes or blows up
to infinity are excluded from the bulk of the spectrum.

\begin{theorem}\label{thm:level-density}
  For the orthogonal, unitary or Jacobi ensembles associated to the
  weight function~\eqref{eq:49}, the global level density is given by
  \begin{equation}
    \label{eq:57}
    \rho(x) = \frac1{\pi\sqrt{1-x^2}}\qquad\text{on $(-1,1)$.}
  \end{equation}
  The limit in~\eqref{eq:41} is attained uniformly on compact subsets
  of $(-1,1)$.
\end{theorem}
This theorem will be proved in the following section.

If we revert to the angular variable $\theta$ with $x=\cos\theta$, we see that
\begin{equation}
  \label{eq:58}
  \rho(x)dx = \frac{d\theta}\pi = \varrho(\theta)d\theta
\end{equation}
so the level density $\varrho(\theta)\equiv1/\pi$ on $(0,\pi)$ is constant: the
eigenvalues become equidistributed on the unit circle (with respect to
its invariant measure), and uniformly so away from the central
eigenvalues $\pm1$, in the semiclassical limit $R\to\infty$.  The bulk of the
spectrum excludes the edges $\pm1$.

The local $n$-level correlations are the ``local'' semi-classical 
limits of the $n$-level correlations $I_R^{(n)}$.  When localizing
near the neighborhood of a fixed level $z_o$ belonging to the bulk of
the spectrum, these local correlations are universal in the sense that
they depend neither on the specific ensemble nor on the choice of
$z_o$ but only on the symmetry parameter $\beta$.  In particular they
coincide with the local correlations of the Gaussian Orthogonal
($\beta=1$), Unitary ($\beta=2$) or Symplectic ($\beta=4$) ensemble,
respectively.  For Jacobi ensembles the bulk of the spectrum consists
of the open interval $(-1,1)$, whereas the local correlations near the
``hard edges'' $\pm1$ (which correspond to the ``central eigenvalues''
$\pm1$ on the unit circle) have a different behavior which is sensitive
to the parameters $(a,b)$ of the ensemble.

\begin{rem}
  As we shall see later, the level density vanishes to some order at,
  say, the hard edge $+1$ depending on the parameter $a$ (which is
  natural since $a$ determines the order to which the weight
  function~\eqref{eq:49} vanishes at $x_j=+1$).  The local
  correlations fail to follow Dyson's universal ``threefold way'', but
  rather depend on this parameter.  The same limiting behavior occurs
  at the hard edge $0$ of Laguerre ensembles~\cite{N-F}, so that, at
  least conjecturally, these ``universal'' laws---manifestly different
  from Dyson's bulk regimes---describe the behavior of the local
  correlations at a hard edge for general orthogonal, unitary or
  symplectic ensembles.
\end{rem}

We now fix a level $z_o\in[-1,1]$.  Given that the eigenvalue density
is uniform, it is natural to change variables from $\bfx$ to $\bfxi$
stretching the angles by a factor $R$, namely setting
\begin{equation}
  \label{eq:59}
  x_j = \cos\left(\alpha_o+\frac\pi R \xi_j\right),
\end{equation}
where $\alpha_o=\arccos z_o$ (note that the change of variables depends on
$R$).  The semiclassical limit of the correlation functions is
obtained by letting $R$ tend to infinity.  What the factor $\pi/R$
accomplishes is that, on the bulk of the spectrum, the local level
density (i.e., the local limit of the correlation function
$I_{R\beta}^{(1)}$) will be $\bar\rho(\xi)\equiv1$.

\begin{theorem}\label{thm:loc-corr}
  For the orthogonal ($\beta=1$), unitary ($\beta=2$) and symplectic
  ($\beta=4$) Jacobi ensembles associated to the weight
  function~\eqref{eq:49}, the local correlations are as follows:
  \begin{itemize}
  \item Bulk local correlations (independent of $\beta$ and of the choice
    of a fixed $z_0=\cos\alpha_0\in(-1,1)$).
      \begin{itemize}
        \item Local level density:
          \begin{equation}
            \label{eq:134}
            \bar\rho(\xi) =
            \lim_{R\to\infty} (R\rho(x))^{-1} I_{R\beta}^{(1)}(x)
            \equiv 1,\qquad\text{$\xi\in\bfR$}.
          \end{equation}
          where $x$ depends on $\xi$ as in~\eqref{eq:59} and $\rho(x)$ is
          the global level density~\eqref{eq:57}.
        \item Local correlations:
          \begin{equation}
            \label{eq:131}
              L_\beta^{(n)}(z_o;\bfxi_n) = \lim_{R\to\infty} (R\rho(z_o))^{-n}
              I_{R\beta}^{(n)}(\bfx_n) = \Det(\bar K_\beta(\xi_j,\xi_k))_{n\times n},
          \end{equation}
          where $\bfx_n$ and $\bfxi_n$ are related by~\eqref{eq:59}
          (recall that $\Det$ stands for the usual or the quaternion
          determinant in the cases of $\beta=2$ and $\beta=1,4$,
          respectively).  In the case $\beta=2$, $\bar K_2$ is the scalar
          Sine Kernel
          \begin{equation}
            \label{eq:60}
            \bar K_2(\xi,\eta) = \begin{cases}
              \frac{\sin\pi(\xi-\eta)}{\pi(\xi-\eta)},&\text{$\xi\neq\eta$;}\\
              \bar\rho(\xi)=1,&\text{$\xi=\eta$}.\end{cases}
          \end{equation}
          In the case $\beta=4$ the matrix Sine Kernel $\bar K_4$ is
          given by
          \begin{equation}
            \label{eq:61}
            \bar K_4(\xi,\eta) =
            \begin{pmatrix}
              \bar S_4(\xi,\eta) & \bar I_4(\xi,\eta) \\
              \bar D_4(\xi,\eta) & \bar S^T_4(\xi,\eta)
            \end{pmatrix},
          \end{equation}
          where
          \begin{eqnarray}
            \label{eq:62}
              \bar S_4(\xi,\eta) &=& \bar K_2(2\xi,2\eta), \\
              \bar I_4(\xi,\eta) &=& -\int_\xi^\eta \bar S_4(\xi,t)dt, \nonumber \\
              \bar D_4(\xi,\eta) &=& \partial_\xi \bar S_4(\xi,\eta), \nonumber \\
              \bar S_4^T(\xi,\eta) &=& \bar S_4(\eta,\xi).\nonumber
            \end{eqnarray}
            In the case $\beta=1$ the matrix Sine Kernel $\bar K_1$ is
            given by
            \begin{equation}
              \label{eq:61a}
              \bar K_1(\xi,\eta) =
              \begin{pmatrix}
                \bar S_1(\xi,\eta) & \bar I_1(\xi,\eta) - \epsilon(\xi-\eta) \\
                \bar D_1(\xi,\eta) & \bar S^T_1(\xi,\eta)
              \end{pmatrix},
            \end{equation}
            where
            \begin{eqnarray}
              \label{eq:62a}
              \bar S_1(\xi,\eta) &=& \bar K_2(\xi,\eta), \\
              \bar I_1(\xi,\eta) &=& -\int_\xi^\eta \bar S_1(\xi,t)dt, \nonumber \\
              \bar D_1(\xi,\eta) &=& \partial_\xi \bar S_1(\xi,\eta), \nonumber \\
              \bar S_1^T(\xi,\eta) &=& \bar S_1(\eta,\xi).\nonumber
            \end{eqnarray}
          \end{itemize}
   \item Hard edge $z_o=+1$ ($\alpha_o=0$).
      \begin{itemize}
        \item Central point level density.  For $\xi>0$:
          \begin{equation}
            \label{eq:136}
            \lim_{R\to\infty} \left(\frac R\pi\right)^{-1}
            I_{R\beta}^{(1)}(x) = \hat\rho_\beta(\xi)
          \end{equation}
          (where $x$ depends on $\xi$ by~\eqref{eq:59}) is given by:
          \begin{eqnarray}
            \label{eq:121}
            \hat\rho_2^{(a)}(\xi) &=& \frac\pi2(\pi\xi)[J_a(\pi\xi)^2 -
            J_{a-1}(\pi\xi)J_{a+1}(\pi\xi)], \\
            \label{eq:69}
            \hat\rho_1^{(a)}(\xi) &=& \hat\rho_2^{(2a+1)}(\xi) + \frac\pi2
            J_{2a+1}(\pi\xi)\int_{\pi\xi}^\infty J_{2a+1}, \\
            \label{eq:70}
            \hat\rho_4^{(a)}(\xi) &=& \hat\rho_2^{(a)}(2\xi)-\frac\pi2
            J_{a-1}(2\pi\xi)\int_0^{2\pi\xi}J_{a+1}.
          \end{eqnarray}
        \item Local correlations.  For $\bfxi_n>0:$
          \begin{equation}
            \label{eq:132}
            L_\beta^{(n)}(+1;\bfxi_n) = \lim_{R\to\infty} \left(\frac R\pi\right)^{-n} 
            I_{R\beta}^{(n)}(\bfx_n) = \Det(\hat K_\beta(\xi_j,\xi_k))_{n\times n},
          \end{equation}
          with $\bfx_n$ related to $\bfxi_n$ by~\eqref{eq:59}.  The
          scalar ``Bessel Kernel'' $\hat K_2=\hat K_2^{(a)}$ is given
          by
          \begin{equation}
            \label{eq:68}
            \hat{K}_2^{(a)}(\xi,\eta) = \begin{cases} \frac{\sqrt{\xi\eta}}{\xi^2-\eta^2}
            [\pi\xi J_{a+1}(\pi\xi)J_a(\pi\eta) - J_a(\pi \xi) \pi\eta
            J_{a+1}(\pi\eta)], & \xi\neq\eta;  \\
            \hat\rho_2^{(a)}(\xi), & \xi=\eta .\end{cases}
          \end{equation}
          For $\beta=1,4$ the matrix Bessel Kernels are given by the same
          expressions of~\eqref{eq:61}--\eqref{eq:62a}, except that
          the bars are to be replaced by hats and $\hat S_1=\hat
          S_1^{(a)},\hat S_4=\hat S_4^{(a)}$ are given by
          \begin{align}
            \label{eq:42}
            \hat{S}^{(a)}_1(\xi,\eta) &=& \sqrt\frac\xi\eta \hat{K}_2^{(2a+1)}(\xi,\eta) +
            \frac\pi2 J_{2a+1}(\pi\eta) \int_{\pi\xi}^\infty J_{2a+1}(t)dt,\\
            \label{eq:63}
            \hat{S}^{(a)}_4(\xi,\eta) &=& \sqrt\frac\xi\eta
            \hat{K}_2^{(a-1)}(2\xi,2\eta) - \frac\pi2
            J_{a-1}(2\pi\eta)\int_0^{2\pi\xi}J_{a-1}(t)dt.
          \end{align}
          where the $J_\nu$ are the Bessel functions of the first kind.
        \end{itemize}
      \end{itemize}
\end{theorem}
The next section will be devoted to the proof of this theorem.
\begin{rem}
  The local limits at the edge $z_0=-1$ are given by the same formulae
  replacing the parameter $a$ by $b$.
\end{rem}
\begin{rem}
  The integral in~\eqref{eq:63} diverges for $-1<a<0$.  However, in
  the next section we provide an alternative version of that equation
  which is well-defined for all $a>-1$.
\end{rem}
\begin{rem}
  In connection with the hard edge correlations for the classical
  orthogonal and symplectic groups (table~\ref{tab:prob-eigen-II}), we
  remark that the unitary Bessel kernel~\eqref{eq:68}, in the case
  $a=+1/2$ (resp., $a=-1/2$), coincides with the ``odd'' (resp.,
  ``even'') Sine Kernel~\cite{K-S}:
  \begin{equation}
    \label{eq:91}
    \bar K_2^{(\pm1/2)}(\xi,\eta) = \frac{\sin(\xi-\eta)}{\xi-\eta} \mp
    \frac{\sin(\xi+\eta)}{\xi+\eta}.
  \end{equation}
\end{rem}
\begin{figure}
\includegraphics[scale=1.4]{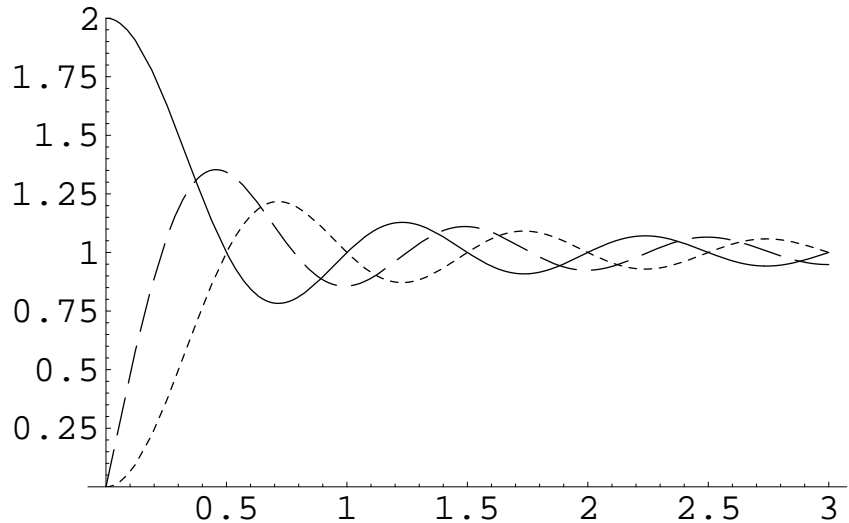}
\caption[$\beta=2$: Level density at the central point]{Graphs of $\hat\rho_2^{(a)}(\xi)$ for $a=-1/2$ (the ``even'' Sine Kernel, solid), $a=+1/2$ (the ``odd'' Sine Kernel, dotted), and $a=0$ (the Legendre Kernel, dashed).}
\end{figure}

\begin{figure}
\includegraphics[scale=1.4]{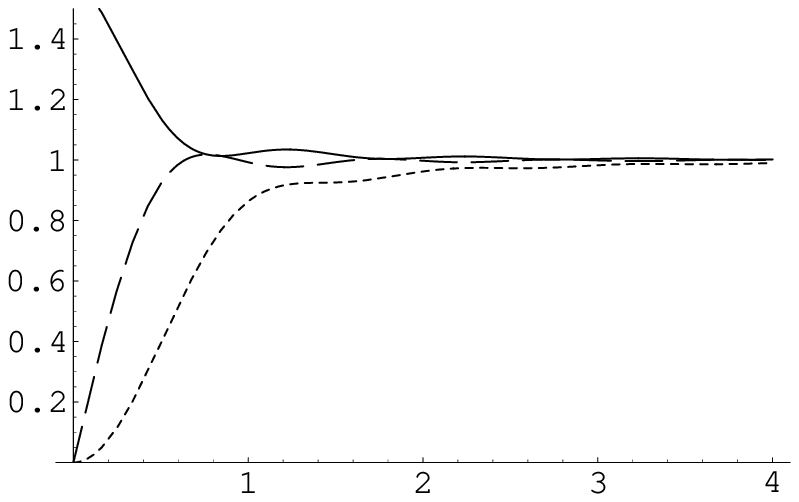}
\caption[$\beta=1$: Level density at the central point]{Graphs of $\hat\rho_1^{(a)}(\xi)$ for $a=-1/2$ (solid), $a=+1/2$ (dotted), and $a=0$ (dashed).}
\end{figure}

\begin{figure}
\includegraphics[scale=1.4]{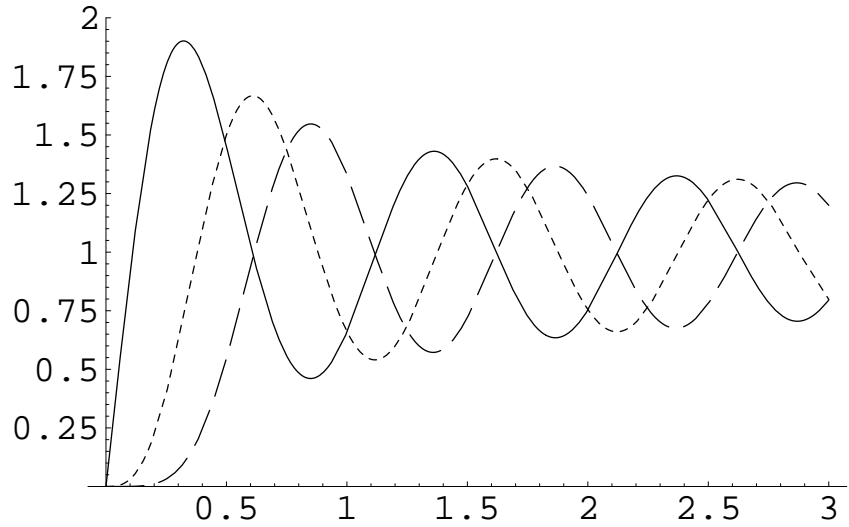}
\caption[$\beta=2$: Level density at the central point]{Graphs of $\hat\rho_4^{(a)}(\xi)$ for $a=0$ (solid), $a=1$ (dotted), and $a=2$ (dashed).}
\end{figure}

\section{Proofs}
\label{sec:proof}
In this section we prove theorems~\ref{thm:level-density}
and~\ref{thm:loc-corr}.  First we remark that the unitary case has
been studied in the work of Nagao and Wadati~\cite{N-W1,N-W2}, but we
reproduce the proofs here for completeness and also to show that the
hypothesis $a>-1$ is, in a certain sense, unnecessary.  Also we remark
that Forrester and Nagao~\cite{N-F} have studied the hard edge
correlations directly, using skew-orthogonal polynomial expressions
for the matrix kernels $K_{R1},K_{R4}$, but their results apply only
when the parameters $a,b$ are strictly positive, and in view of the
application to symmetric spaces this restriction is unacceptable (see
table~\ref{tab:prob-eigen}).  Also, their somewhat more complicated
formulas for the limiting quantities $\hat S_1,\hat S_4$ are given in
terms of iterated integrals of Bessel functions.  Here we take
advantage of the more recent work of Adler~\emph{et al} which provides
simple ``summation formulas'' for the quantities $S_{R1},S_{R4}$.

\subsection{Some preliminary results and formulas}
\label{sec:prelim}
The various results we quote on Jacobi polynomials can be found in
Szeg\H{o}'s book~\cite{Sze-op} and in his article on asymptotic
properties of Jacobi polynomials~\cite{Sze-jp} (reproduced in his
collected papers~\cite{Sze-cp2}).  Stirling's formula and the Bessel
function identities can be found, for instance, in the tables of
Gradshteyn and Ryzhik~\cite{MR33:5952}.  We denote by $\pAB_N(x)$ the
classical Jacobi polynomials defined by
\begin{equation}
  \label{eq:jac-pol}
  (1-x)^A(1+x)^B\pAB_N(x) =
  \frac{(-1)^N}{2^NN!}\left(\frac d{dx}\right)^N\left[(1-x)^{N+A}(1+x)^{N+B}\right].
\end{equation}
When $A,B>-1$, these polynomials are orthogonal on $[-1,1]$ with
respect to the weight
\begin{equation}
  \label{eq:71}
  w(x)=|1-x|^A|1+x|^B,
\end{equation}
but they are not normalized.  However, the formula~\eqref{eq:jac-pol}
is meaningful for arbitrary (real or complex) values of the parameters
$A,B$, and defines a polynomial in $A,B,x$ of degree (at most) $N$ in
$x$.  In fact
\begin{equation}
  \label{eq:150}
  \pAB_N(x) = \sum_{k=0}^N \binom{A+N}{k} \binom{B+N}{N-k}
  \left(\frac{x-1}2\right)^{N-k}\left(\frac{x+1}2\right)^k.
\end{equation}
In particular
\begin{equation}
  \label{eq:151}
  \pAB_N(+1) = \binom{A+N}{N}.
\end{equation}
The derivative of a Jacobi polynomial is related to another Jacobi
polynomial by the identity (the apostrophe denotes differentiation
with respect to $x$)
\begin{equation}
  \label{eq:126}
  {\pAB_N}'(x) = \frac12(N+A+B+1)P^{(A+1,B+1)}_{N-1}(x).
\end{equation}

\begin{proposition}[Darboux's formula]
  (With an improved error term due to Szeg\H{o}~\cite{Sze-jp}.)  For
  arbitrary reals $A,B$,
  \begin{gather}
    \begin{split}
    \label{eq:darboux}
    \pAB_N(\cos\theta) &= (\pi N)^{-1/2}\left(\sin\frac\theta2\right)^{-A-1/2}
    \left(\cos\frac\theta2\right)^{-B-1/2} \cos(N'\theta+\gamma)\\
    & \quad + E,      
    \end{split}\\
    N' = N+\frac{A+B+1}2,\qquad \gamma=-\left(A + \frac12\right)\frac\pi2,
  \nonumber
  \end{gather}
  for $0<\theta<\pi$, where the error term $E$ satisfies
  \begin{equation}
    \label{eq:123}
    E = \theta^{-A-3/2}O(N^{-3/2}),\qquad\text{uniformly for $c/N\leq\theta\leq\pi-\epsilon$,}
  \end{equation}
for any positive constants $c,\epsilon$, and the constant implied by the $O$
symbol depends only on $c,\epsilon,A,B$.  
\end{proposition}

\begin{proposition}[Hilb's formula](As generalized by Szeg\H{o} to Jacobi
  polynomials~\cite{Sze-op}.)  For $A>-1$ and any real $B$:
  \begin{equation}
    \begin{split}
      \label{eq:hilb1}
      \left(\sin\frac\theta2\right)^A\left(\cos\frac\theta2\right)^B
      \pAB_N(\cos\theta)
      &= N^{-A}\frac{\G(N+A+1)}{N!}\sqrt{\frac\theta{\sin\theta}}J_A(N'\theta)\\
      & \quad + E,    
    \end{split}
  \end{equation}
where $N'$ has the same meaning as in~(\ref{eq:darboux}) and the error
term $E$ is given by
\begin{equation}
  \label{eq:err}
  E=
  \begin{cases}
    \theta^{1/2}O(N^{-3/2})&\text{if $c/N\leq\theta\leq\pi-\epsilon$},\\
    \theta^{A+2}O(N^A)&\text{if $0<\theta\leq c/N$},
  \end{cases}
\end{equation}
where $c,\epsilon$ are arbitrary but fixed positive constants, and the
constants implied by the $O$ symbol depend on $A,B,c,\epsilon$ only.
\end{proposition}

The restriction to $A>-1$, however, is too strong for some purposes,
and we will need the following formula, also due to
Szeg\H{o}~\cite{Sze-jp} (reproduced in~\cite{Sze-cp2}):
\begin{equation}
  \begin{split}
    \label{eq:hilb2}
    \pAB_N(\cos\theta) &=
    \left(\sin\frac\theta2\right)^{-A}\left(\cos\frac\theta2\right)^{-B}
    \sqrt{\frac\theta{\sin\theta}}\left(1-\sqrt{\frac{\tan(\theta/2)}{2\theta}}\right)\times\\
    & \quad \times J_A(N'\theta) + R,
  \end{split}
\end{equation}
with $N'$ as in~\eqref{eq:darboux}.  Here $A,B$ are arbitrary reals.
The error term $R$ satisfies:
\begin{equation}
  \label{eq:err-hilb2}
  R =
  \begin{cases}
    \theta^{\frac12-A}O(N^{-3/2}) & \mbox{if $c/N\leq\theta\leq\pi-\epsilon$,}\\
    O(N^{A-2}) & \mbox{if $0<\theta\leq c/N$},
  \end{cases}
\end{equation}
where $c,\epsilon$ are fixed positive numbers, and the constants implied by
the $O$ symbol depend only on $A,B,c,\epsilon$.  It must be noted, however,
that the error term $R$ of~(\ref{eq:err-hilb2}) does not depend on $\theta$
on the range $0<\theta<c/N$, which makes this formula less useful
than~\eqref{eq:hilb1} with the error term~\eqref{eq:err} for $\theta$ in
this range.

Recall Stirling's asymptotic formula for the Gamma function:
\begin{equation}
  \label{eq:stir}
  \log\G(x) = \left(x-\frac12\right)\log x - x + \frac12\log2\pi +
  O(x^{-1}),
  \qquad\text{as $x\to\infty$.}
\end{equation}

The Bessel functions of the first kind are defined by the series
\begin{equation}
  \label{eq:120}
  J_\nu(z) = \left(\frac z2\right)^\nu \sum_{k=0}^\infty
  (-1)^k\frac{z^{2k}}{2^{2k}k!\G(\nu+k+1)}, \quad z\in\bfC\backslash(-\infty,0],\quad\nu\in\bfR;
\end{equation}
they satisfy, among many others, the relations:
\begin{eqnarray}
  \label{eq:121a}
  J_\nu'(z) &=& J_{\nu-1}(z) - \frac\nu z J_\nu(z), \\
  \label{eq:121b}
  J_\nu'(z) &=& -J_{\nu+1}(z) + \frac\nu z J_\nu(z), \\
  \label{eq:121c}
  J_\nu'(z) &=& \frac12[J_{\nu-1}(z)-J_{\nu+1}(z)], \\
  \label{eq:121d}
  J_{\nu+1}(z) &=& \frac{2\nu}z J_\nu(z) - J_{\nu-1}(z), \\
  \label{eq:121e}
  \frac d{dz}[z^\nu J_\nu(z)] &=& z^\nu J_{\nu-1}(z), \\
  \label{eq:121f}
  \frac d{dz}[z^{-\nu} J_\nu(z)] &=& -z^{-\nu} J_{\nu+1}(z).
\end{eqnarray}
We also have
\begin{eqnarray}
  \label{eq:157}
  \int J_\nu &=& 2\sum_{k=0}^\infty J_{\nu+2k+1}, \\
  \label{eq:149}
  \int_0^\infty J_\nu(t)dt &=& 1 \qquad\text{for $\nu>-1$}.
\end{eqnarray}

\subsection{Asymptotics of the Unitary Jacobi Kernel}\label{sec:asympt-unit-jacobi}
In this section we recall the proofs of some of the results of Nagao
and Wadati~\cite{N-W1}, which will be needed later on in the analysis
of the orthogonal and symplectic cases.

Using the Christoffel-Darboux summation formula~\cite{Sze-op}, the
scalar kernel $\KAB_{N2}$ can be written in the form
\begin{equation}
  \label{eq:118}
  \begin{split}
    K^{(A,B)}_{N2}(x,y)
    &= \frac{2^{-A-B}}{2N+A+B}\frac{\G(N+1)\G(N+A+B+1)}{\G(N+A)\G(N+B)}\\
    & \quad
    \times\sqrt{w(x)w(y)}\frac{\pAB_{N}(x)\pAB_{N-1}(y)-\pAB_{N-1}(x)\pAB_N(y)}{x-y},
  \end{split}
\end{equation}
for $x\neq y$, and
\begin{equation}
  \label{eq:124}
  \begin{split}
    K^{(A,B)}_{N2}(x,x)
    & =
    \frac{2^{-A-B}}{2N+A+B}\frac{\G(N+1)\G(N+A+B+1)}{\G(N+A)\G(N+B)} \\
    & \quad \times w(x)[{\pAB_{N}}'(x)\pAB_{N-1}(x) - {\pAB_{N-1}}'(x)\pAB_N(x)].
  \end{split}
\end{equation}
We observe that the kernel $K_{N2}$ given by~\eqref{eq:118}
and~\eqref{eq:124} is well-defined for $A,B>-c$ for any real constant
$c$ provided $N$ is sufficiently large.

First consider the global level density
\begin{equation}
  \label{eq:72}
  \rho(x) = \lim_{N\to\infty}N^{-1}K(x,x).
\end{equation}
Using Darboux's formula~(\ref{eq:darboux}) together with the
identity~(\ref{eq:126}) in the expression~\eqref{eq:124} for the
kernel, we find:
\begin{equation}
  \label{eq:1220}
   K_{N2}^{(a,b)}(x,x) = \frac N{\pi\sqrt{1-x^2}} + O(1)
\end{equation}
where the implied constant depends only on $\epsilon$ for $-1+\epsilon\leq x\leq1-\epsilon$.
Equation~(\ref{eq:1220}) proves~\eqref{eq:57} (in the unitary case).

A density function $D=D(x_1,\dots,x_n)$ defines a measure
$D\,dx_1\dots dx_n$.  Under a (monotonically increasing or decreasing)
differentiable change of variables $x_j=X(u_j)$, this density is
transformed into the density
\begin{equation}
  \label{eq:138}
  \mathcal{D}(u_1,\dots,u_n) =\left( \prod_{j=1}^n|X'(u_j)|\right) D(X(u_1),\dots,X(u_n)).
\end{equation}
If the density $D$ is given as a determinant with a (scalar) kernel
$K(x,y)$, namely $D=\det(K(x_j,x_k))_{n\times n}$, then the change of
variables reflects itself in the kernel in the following fashion:
\begin{lemma} \label{lem:resc2}
  After the (monotonic) differentiable change of variables $u\to
  x=X(u)$, the correlation functions are given as the
  determinant~\eqref{eq:51} defined using the kernel
  \begin{equation}
    \label{eq:142}
    \mathcal{K}(u,v) = \sqrt{|X'(u)X'(v)|}K(X(u),X(v)).
  \end{equation}
\end{lemma}
This is clear since the introduction of the factor
$\sqrt{|X'(u)X'(v)|}$ results in multiplying the
determinant~\eqref{eq:51} by $\prod_{j=1}^n|X'(u_j)|$.

The localization at some $-1<z_o=\cos\alpha_o<1$ given by the change of
variables~\eqref{eq:59} leads us to consider the limit
\begin{equation}
  \label{eq:125}
  \begin{split}
    \bar K_2^{(a,b)}(\xi,\eta)
    & = \lim_{N\to\infty} \big(N\sqrt{\rho(x)\rho(y)}\big)^{-1} K_{N2}^{(a,b)}(x,y)\\
    & = \lim_{N\to\infty} (N\rho(z_o))^{-1} K_{N2}^{(a,b)}(x,y),
  \end{split}
\end{equation}
with $x,y$ related to $\xi,\eta$ by~\eqref{eq:59}, which from Darboux's
formula~\eqref{eq:darboux} can be easily seen to be the Sine
Kernel~\eqref{eq:60}, independently of the value of $z_o$ (as long as
$-1<z_o<1$), for any real $a,b$, and the limit is attained uniformly
on compacta.

For the localization at $z_o=+1$ ($\alpha_o=0$) ---localization at
$z_o=-1$ is analogous provided $a$ and $b$ are interchanged---, we use
the same change of variables~\eqref{eq:59} with $\bfxi_n>0$.  To
compute the limit
\begin{equation}
  \label{eq:127}
  \begin{split}
    \hat K_2^{(a,b)}(\xi,\eta) 
    &= \lim_{N\to\infty} \big(N\sqrt{\rho(x)\rho(y)}\big)^{-1} K_{N2}^{(a,b)}(x,y) \\
    &= \lim_{N\to\infty} (N\rho(z_o))^{-1} K_{N2}^{(a,b)}(x,y),
  \end{split}
\end{equation}
we use Szeg\H{o}'s formulas~(\ref{eq:hilb1}), (\ref{eq:hilb2}), in
conjunction with~(\ref{eq:118}) and~(\ref{eq:124}):
\begin{equation}
  \label{eq:155}
  \hat K_2^{(a)}(\xi,\eta) = \frac{\sqrt{\xi\eta}}{\xi^2-\eta^2}
  [\pi\xi J_a'(\pi\xi)J_a(\pi\eta) - J_a(\pi \xi) \pi\eta J_a'(\pi \eta)].
\end{equation}
Using the derivation formula~(\ref{eq:121b}) we rewrite this kernel in
the form~\eqref{eq:68}.  For the case $\xi=\eta$ we start with the
expression~(\ref{eq:124}) and use the derivation
formula~(\ref{eq:126}) to find:
\begin{equation}
  \label{eq:165}
  \begin{split}
    \hat\rho_2^{(a)}(\xi)&=\hat K_2^{(a)}(\xi,\xi) \\
    &=
    \frac\pi2[J_a(\pi\xi)J_{a+1}(\pi\xi) + \pi\xi J_{a+1}'(\pi\xi)J_a(\pi\xi)-\pi\xi J_a(\pi\xi)'J_{a+1}(\pi\xi)].
  \end{split}
\end{equation}
Applying the derivation formula~(\ref{eq:121a}) and the recurrence
formula~(\ref{eq:121d}) this can be rewritten in the
form~\eqref{eq:121}.

\subsection{Asymptotics of the Orthogonal Jacobi Kernel}
\label{sec:orth-local-limit}

We start with some general remarks.  If a density $P=P(x_1,\dots,x_n)$
is given as a quaternion determinant with a self-dual matrix kernel
$K(y,x)=K(x,y)^D$, namely $P=\qdet(Q(x_j,x_k))_1^n$, then under a
differentiable change of variables $x_j=X(u_j)$ the density is still
given as a quaternion determinant.
\begin{lemma} \label{lem:ch-var}
  After a (monotonic) differentiable change of variables $u\to
  x=X(u)$, a density function
  \begin{equation}
    \label{eq:143}
    P(x_1,\dots,x_n) = \qdet(K(x_j,x_k))
  \end{equation}
  defined in terms of some self-dual matrix kernel ($\delta=0,1$)
  \begin{equation}
    \label{eq:74}
    K(x,y) =
    \begin{pmatrix}
      S(x,y) & I(x,y)-\delta\epsilon(x-y) \\
      D(x,y) & S^T(x,y)
    \end{pmatrix}
  \end{equation}
  with
  \begin{align}
    \label{eq:75}
    I(x,y) &= -\int_x^y S(x,z)dz, \\
    \label{eq:81}
    D(x,y) &= \partial_x S(x,y),  \\
    \label{eq:82}
    S^T(x,y) &= S(y,x).
  \end{align}
  is transformed into the density
  \begin{equation}
    \label{eq:144}
    \mathcal{P}(u_1,\dots,u_n) = \qdet(\mathcal{K}(u_j,u_k)),
  \end{equation}
  where
  \begin{align}
    \label{eq:76}
    \mathcal{K}(u,v) &=
    \begin{pmatrix}
      \mathcal{S}(u,v) & \mathcal{I}(u,v)-\delta\epsilon(u-v) \\
      \mathcal{D}(u,v) & \mathcal{S}^T(u,v)
    \end{pmatrix}\\
    \label{eq:77}
    \mathcal{S}(u,v) &= S(X(u),X(v))|X'(v)| = \pm S(X(u),X(v))X'(v) \\
    \label{eq:78}
    \mathcal{I}(u,v) &= -\int_u^v \mathcal{S}(u,w)dw, \\
    \label{eq:79}
    \mathcal{D}(u,v) &= \partial_u\mathcal{S}(u,v), \\
    \label{eq:80}
    \mathcal{S}^T(u,v) &= \mathcal{S}(v,u).
  \end{align}
\end{lemma}
For the proof, we need first:
\begin{lemma} \label{lem:sandwich}
  Let $H = H^D = J_nH^TJ_n^T$ be a $2n\times2n$ self-dual complex matrix.
  Let $k_j$, $j=1,2,\dots,n$ be arbitrary complex constants.  Set
  $K=\diag(k_1,\dots,k_n)$.  Then the matrices
  \begin{equation}
    \label{eq:139}
    H_1 = \diag(I,K)H\diag(K,I)\qquad H_2=\diag(-I,K)H\diag(-K,I)
  \end{equation}
  (where $I=I_n$ is the $n\times n$ identity matrix) are both self-dual,
  and
  \begin{equation}
    \label{eq:140}
    \qdet(H_1) = \det(K)\qdet(H) = \qdet(H_2).
  \end{equation}
\end{lemma}
The verification that $H_1$ and $H_2$ are self-dual is trivial.  On
the other hand, since $(\qdet X)^2=\det X$ for any self-dual matrix
$X$, we have that
\begin{equation}
  \begin{split}
    \label{eq:141}
    (\qdet(H_1))^2 &= (\qdet(H_2))^2 = (\det(K))^2\det(H) \\
    & = (\det(K))^2(\qdet(H))^2.
  \end{split}
\end{equation}
Hence equation~(\ref{eq:140}), which is an equality between
polynomials in the entries of the matrices involved, must hold up to a
sign.  Setting $K=I_n$ we see that the first equality
in~\eqref{eq:140} holds, and setting $K=-I_n$, so $H_2=-H$, the
validity of the second equality in~\eqref{eq:140} is equivalent to the
easy fact that $\qdet(-H)=(-1)^n\qdet H=\det(-I_n)\qdet H$.

Proceeding to the proof of lemma~\ref{lem:ch-var}, we first observe
that, after the change of variables $u\to x$, the density
$P(x_1,\dots,x_n)$ transforms into the density
\begin{equation}
  \label{eq:83}
  \mathcal{P}(u_1,\dots,u_n) = P(X(u_1),\dots,X(u_n))\prod_{j=1}^n |X'(u_j)|.
\end{equation}
We apply lemma~\ref{lem:sandwich} with $H=(K(X(u_j),X(u_k)))_{n\times n}$
and $k_j=|X'(u_j)|$ to conclude that~\eqref{eq:144} holds with either
of the two kernels (we write $X(u,v)$ for $(X(u),X(v))$)
\begin{equation}
  \label{eq:84}
  \mathcal{K}_\pm(u,v) = 
  \begin{pmatrix}
    S(X(u,v))|X'(v)| & \pm(I-\delta\epsilon)(X(u,v)) \\
    \pm D(X(u,v))|X'(u)||X'(v)| & S^T(X(u,v))|X'(u)|
  \end{pmatrix}.
\end{equation}
The plus and minus signs correspond to applying the first and second
of the equalities in~\eqref{eq:140}, respectively.  If $x\to u$
preserves orientation, then we observe that $\epsilon(X(u)-X(v))=\epsilon(u-v)$
and conclude by a simple application of the chain rule and a change of
variables in the integral that the kernel $\mathcal{K}_+$ coincides
with $\mathcal{K}$ from~\eqref{eq:76} for the
choices~\eqref{eq:77}--\eqref{eq:80}.  If $x\to u$ reverses
orientation, we choose the minus signs, observe that
$\epsilon(X(u)-X(v))=-\epsilon(u-v)$ and proceed exactly as before to see that
$\mathcal{K}_-$ coincides with~\eqref{eq:76} in this case.

Lemma~\ref{lem:ch-var} explains the relations~\eqref{eq:62a} between
the entries of the limiting kernels $\bar K_\beta$ and also of $\hat
K_\beta$ ($\beta=1,4$).  The relations certainly hold when $R$ is finite
after applying the change of variables~\eqref{eq:59} to the the matrix
kernel $K_{R\beta}$ so as to obtain another kernel $\CK_{R\beta}$.  They can
be shown to continue to hold in the limit either by noting that the
sequence of scalar kernels $\{\CS_{R\beta}(\xi,\eta)\}_{R=0}^\infty$ is a normal
sequence of analytic functions (i.e., it converges uniformly on
compacta), or by direct verification that each of the sequences
$\{\CS_{R\beta}\},\{\CI_{R\beta}\},\{\CK_{R\beta}\},\{\CS^T_{R\beta}\}$ converges
to the correct limit as $R\to\infty$.  In what follows we will only
consider the limit of the quantity $S_{R\beta}$ which alone determines
the matrix kernel $K_{R\beta}$.

Let $A=2a+1, B=2b+1$, where $a,b$ are the parameters of the orthogonal
Jacobi ensemble.  Assume also that $R$ is even. Observe that $A,B>-1$
if $a,b>-1$.  The summation formula of Adler \emph{et al}~\cite{A-}
expresses the orthogonal kernel $S^{(a,b)}_{R1}$ using the unitary
kernel $K^{(A,B)}_{R-1,2}$ and another term.  As we shall see, this
other term is negligible in the localized limit (in the bulk of the
spectrum), but it \emph{does} contribute to the edge limit.

The summation formula for the quantity $S^{(a,b)}_{R1}(x,y)$
of~\eqref{eq:52} is as follows~\cite{A-}:
\begin{equation}
  \label{eq:sum-1}
  S^{(a,b)}_{R1}(x,y) =
  \sqrt{\frac{1-x^2}{1-y^2}}K^{(A,B)}_{R-1,2}(x,y) + c_{R-2} \psi_{R-1}(y)\epsilon\psi_{R-2}(x).
\end{equation}
Here $\epsilon$ denotes the integral operator (cf., eq.~\eqref{eq:53})
\begin{equation}
  \label{eq:73}
  (\epsilon f)(x) = \int_{-1}^1\epsilon(x-y)f(y)dy,
\end{equation}
and we have set
\begin{equation}
  \label{eq:psi-1}
  \psi_N(t) = \psi_N^{(A,B)}(t) = (1-t)^{(A-1)/2}(1+t)^{(B-1)/2}\pAB_N(t)
\end{equation}
and
\begin{equation}
  \label{eq:c-n}
  c_N = 2^{-A-B-1} \frac{\G(N+2)\G(N+A+B+2)}{\G(N+A+1)\G(N+B+1)}.
\end{equation}
The quantity $\Sab_{R1}$ determines the entries of the matrix kernel
$\Kab_{R1}$ as per equations~\eqref{eq:85}--\eqref{eq:88}.

From Stirling's formula~(\ref{eq:stir}), the asymptotic behavior of
the coefficient $c_N$ is
\begin{equation}
  \label{eq:cn-inf}
  c_N \sim 2^{-A-B-1}N^2,\qquad\mbox{as $N\to\infty$.}
\end{equation}
\begin{lemma}
  For any real $A,B$:
  \begin{equation}
    \label{eq:166}
    \lim_{N\to\infty}\psi_N^{(A,B)}(\cos\phi) = 0
  \end{equation}
  for $0<\phi<\pi$, uniformly on compacta.
\end{lemma}
This follows immediately from Darboux's formula~(\ref{eq:darboux}).

This lemma is, however, insufficient to understand the asymptotics of
the function $\epsilon\psi_N$ as $N\to\infty$ since it says nothing about the behavior
of $\psi_N$ near the edge.  First we note:
\begin{lemma} \label{lem:psi}
  For $A>-1$ and $B$ arbitrary:
  \begin{eqnarray}
    \label{eq:167a}
    \lim_{N\to\infty}N^{-1}\psi_N^{(A,B)}(\cos(\phi/N)) &=&
    2^{\frac{A+B}2}\frac{J_A(\phi)}\phi, \\
    \label{eq:167b}
    \lim_{N\to\infty}\psi_N^{(A,B)}(\cos(\phi/N))\sin(\phi/N) &=& 2^{\frac{A+B}2} J_A(\phi).
  \end{eqnarray}
  The limits hold uniformly on compact subsets of $(0,\infty)$.
\end{lemma}
These follow from Szeg\H{o}'s formula~(\ref{eq:hilb1}).  
\begin{lemma} \label{lem:epsi}
For $A,B$ real with $A>-1$ and any $0<\theta<\pi$ we have:
  \begin{eqnarray}
    \label{eq:168a}
    \lim_{N\to\infty} N\int_0^\theta \psi_N^{(A,B)}(\cos\phi)\sin\phi\,d\phi &=& 2^{\frac{A+B}2}, \\
    \label{eq:168b}
    \lim_{N\to\infty} N\int_0^{\theta/N} \psi_N^{(A,B)}(\cos\phi)\sin\phi\,d\phi &=&
    2^{\frac{A+B}2}\int_0^\theta J_A.
  \end{eqnarray}
\end{lemma}
These follow again from Szeg\H{o}'s formula~(\ref{eq:hilb1}) and
equation~(\ref{eq:149}).  When $-1<A<0$, the dependence on $\theta$ of the
second of the error terms in~(\ref{eq:err}) is critical to ensure that
the contribution of this error term to the integral is negligible (in
particular, this lemma cannot be proven using the alternate
formula~(\ref{eq:err-hilb2}) unless $A>0$.)
\begin{corollary} \label{cor:epsi}
  For $-1<A,B$ and $0<\theta<\pi$:
  \begin{align}
    \label{eq:156a}
    \lim_{N\to\infty} N(\epsilon\psi_N^{(A,B)})(\cos\theta) &= 0, \\
    \label{eq:156b}
    \lim_{N\to\infty} N(\epsilon\psi_N^{(A,B)})(\cos(\theta/N)) &=
    2^{\frac{A+B}2}\left(1-\int_0^\theta J_A\right) = 2^{\frac{A+B}2}\int_\theta^\infty J_A.
  \end{align}
\end{corollary}
This follows from the previous lemma applied to both $\psi_N^{(A,B)}$ and
$\psi_N^{(B,A)}$.  We also used~(\ref{eq:149}) to obtain the last
equality.

We localize at some $z_o=\cos\alpha_o\in(-1,1)$ using the change of variable
$x\to\xi$ of~\eqref{eq:59}.  The limit to consider is
\begin{equation}
  \label{eq:128}
   \bar S_1^{(a,b)}(\xi,\eta) = \lim_{R\to\infty} (N\rho(y))^{-1}S_{R1}^{(a,b)}(x,y) 
   = \lim_{R\to\infty} (N\rho(z_o))^{-1}S_{R1}^{(a,b)}(x,y)
\end{equation}

By the lemmas above, the second term on the right-hand side
of~\eqref{eq:sum-1} is negligible in the limit.  Also, the factor
$\sqrt{\frac{1-x^2}{1-y^2}}$ is $1$ in the limit.  Thus, the
limit~(\ref{eq:128}) is equal to the limiting unitary kernel, namely
the Sine Kernel, whence the expression~\eqref{eq:62a}.

As for the central point, let us now localize at $z=+1$.  Using the
summation formula~(\ref{eq:sum-1}), lemma~\ref{lem:psi} and
corollary~\ref{cor:epsi}, we readily find:
\begin{equation}
  \begin{split}
    \label{eq:hs1}
    \hat{S}_1^{a}(\xi,\eta) &= \sqrt\frac\xi\eta \hat{K}_2^{(2a+1)}(\xi,\eta) +
    \frac\pi2 J_{2a+1}(\pi\eta)\left[1 - \int_0^{\pi\xi}J_{2a+1}(t)dt\right] \\
    &= \sqrt\frac\xi\eta \hat{K}_2^{(2a+1)}(\xi,\eta) +
    \frac\pi2 J_{2a+1}(\pi\eta) \int_{\pi\xi}^\infty J_{2a+1}(t)dt.
  \end{split}
\end{equation}
As we remarked already, the conditions $a>-1$ and $A>-1$ are
equivalent since $A=2a+1$.  Thus we have derived a weak universality
law for the local correlations at the central points $\pm1$ for any
$a,b>-1$.

\begin{lemma} \label{lem:kap}
  Let $\kappa_\alpha(x,y)=xJ_{\alpha+1/2}(x)J_{\alpha-1/2}(y)-J_{\alpha-1/2}(x)yJ_{\alpha+1/2}(y)$.  Then
  \begin{multline}
    \label{eq:158}
    \sqrt\frac xy \kappa_{\alpha\pm1/2}(x,y) - \sqrt{\frac yx}\kappa_{\alpha\mp1/2}(x,y) \\
    = \mp\left(\frac{x^2-y^2}{\sqrt{xy}}\right)J_{\alpha-1/2\mp1/2}(x)J_{\alpha-1/2\pm1/2}(y).
  \end{multline}
  (This equation stands for two different equations, one with the top
  signs and another with the bottom signs.)
\end{lemma}
We prove the equation with the choice of the top signs (the other case
is analogous).  Indeed, expanding the left-hand side we obtain:
\begin{multline}
  \label{eq:159}
  x^{3/2}J_{\alpha+1}(x)y^{-1/2}J_\alpha(y) - x^{1/2}J_\alpha(x)y^{1/2}J_{\alpha+1}(y) \\
  - x^{1/2}J_\alpha(x)y^{1/2}J_{\alpha-1}(y) + x^{-1/2}J_{\alpha-1}(x)y^{3/2}J_{\alpha}(y).
\end{multline}
The central terms can be combined into $-2\alpha
x^{1/2}J_\alpha(x)y^{-1/2}J_\alpha(y)$ using the identity~(\ref{eq:121d}) and
expanded using this same identity into
$-x^{3/2}J_{\alpha-1}(x)y^{-1/2}J_\alpha(y) - x^{3/2}J_{\alpha+1}(x)y^{-1/2}J_\alpha(y)$.
Two terms cancel out, and the remaining two factor to give the
right-hand side of~(\ref{eq:158}).

We now have, using lemma~\ref{lem:kap},
\begin{equation}
  \begin{split}
    \label{eq:160}
    \sqrt\frac\xi\eta \hat K_2^{(A)}(\xi,\eta) &= \frac{\sqrt{\xi\eta}}{\xi^2-\eta^2}\kappa_{A+1/2}(\pi\xi,\pi\eta) \\
    &= \frac\eta\xi \frac{\sqrt{\xi\eta}}{\xi^2-\eta^2}\kappa_{A-1/2}(\pi\xi,\pi\eta) + \pi J_A(\pi\xi)J_{A-1}(\pi\eta) \\
    &= \sqrt\frac\eta\xi \hat K_2^{(A-1=2a)}(\xi,\eta) - \pi
    J_{A-1}(\pi\xi)J_A(\pi\eta),
  \end{split}
\end{equation}
and similarly
\begin{equation}
  \label{eq:152}
  \sqrt\frac\xi\eta \hat K_2^{(A)}(\xi,\eta)
  = \sqrt\frac\eta\xi \hat K_2^{(A+1)}(\xi,\eta) + \pi J_{A+1}(\pi\xi)J_A(\pi\eta).
\end{equation}
From~(\ref{eq:157}):
\begin{eqnarray}
  \label{eq:161}
  \left(\int_0^{\pi\xi} J_A\right) \pm 2J_{A\mp1}(\pi\xi) =  \int_0^{\pi\xi} J_{A\mp2}.
\end{eqnarray}
The last two equations provide alternative forms of the kernel $\hat
S_{1}^{(a)}$, namely
\begin{align}
  \label{eq:162a}
  \hat S_{1}^{(a)}(\xi,\eta) &= \sqrt\frac\eta\xi \hat{K}_2^{(2a)}(\xi,\eta) +
  \frac\pi2 J_{2a+1}(\pi\eta)\left[1 - \int_0^{\pi\xi}J_{2a-1}(t)dt\right] \\
  \label{eq:162b}
  \hat S_{1}^{(a)}(\xi,\eta) &= \sqrt\frac\eta\xi \hat{K}_2^{(2a+2)}(\xi,\eta) +
  \frac\pi2 J_{2a+1}(\pi\eta)\left[1 - \int_0^{\pi\xi}J_{2a+3}(t)dt\right]
\end{align}
As before, the terms in brackets can be replaced by
$\left[\int_{\pi\xi}^\infty\right]$.

\subsection{Asymptotics of the Symplectic Jacobi Kernel}
\label{sec:sympl-local-limit}

Here we set $A=a-1,B=b-1$ where $a,b$ are the parameters of the
symplectic Jacobi ensemble.  Note that here $a,b>-1$ corresponds to
$A,B>-2$.  With $c_N$ as in~\eqref{eq:c-n} and $\psi_N=\psi_N^{(A,B)}$ as
in~(\ref{eq:psi-1}), the summation formula in this case reads
\begin{equation}
  \label{eq:sum-4}
  S_{R4}^{(a,b)}(x,y) =
  \frac12\sqrt\frac{1-x^2}{1-y^2}K_{2R,2}^{(A,B)}(x,y) - \frac12c_{2R-1}\psi_{2R}(y)\delta\psi_{2R-1}(x),
\end{equation}
where the operator $\delta$ acts by
\begin{equation}
  \label{eq:delta}
  \delta f(x) = \int_x^1f(t)dt.
\end{equation}
The formula~\eqref{eq:sum-4} only holds verbatim when $a>0$ (that is,
$A,B>-1$), since the integral defining $\delta\psi_N^{(A,B)}$ is divergent
for $A\leq-1$.  However, we note that the skew orthogonal polynomials of
the second kind are analytic functions of the parameters $a,b>-1$
(corresponding to $A,B>-2$), hence the kernel $K_{N4}$ is an analytic
function on $a,b>-1$.  Thus, we must find a suitable analytic
continuation of~(\ref{eq:sum-4}) valid for $A,B>-2$.  First we remark
that, although the original kernel $K_{2R,2}^{(A,B)}$ of unitary
Jacobi ensembles is defined for $A,B>-1$, equation~\eqref{eq:118} is
well-defined and analytic for $A,B>-2$ if $R>1$ (which we will
assume).  We write
\begin{multline}
  \label{eq:156}
  \delta\psi^{(A,B)}_N(x) = \int_x^1(1-t)^{(A-1)/2}(1+t)^{(B-1)/2}\pAB_N(t)dt \\
  = \int_x^1(1-t)^{(A-1)/2}(1+t)^{(B-1)/2}(\pAB_N(t)-\pAB_N(1))dt \\
  + \pAB_N(1)\int_x^1 (1-t)^{(A-1)/2}(1+t)^{(B-1)/2}dt.
\end{multline}
The first integral on the right-hand side is well-defined and analytic
for $A>-2$. The term $\pAB_N(1)=\binom{A+N}{N}$ (cf.,
equation~(\ref{eq:151})) vanishes for $A=-1$, which is sufficient to
extend the second integral on the right-hand side to a well-defined
analytic function on the range $A>-2$.  It is easy to rewrite that
integral as an incomplete Beta function and use well-known results to
achieve the extension, but one can also proceed elementarily as
follows.  Integrating the second integral by parts we obtain, for
$A>-1$:
\begin{multline}
  \label{eq:153}
  \binom{A+N}{N}\int_x^1 (1-t)^{(A-1)/2}(1+t)^{(B-1)/2}dt \\
  = \frac2{A+1}\binom{A+N}{N} (1-x)^{(A+1)/2}(1+x)^{(B-1)/2}  \\
    + \frac{B-1}{A+1}\binom{A+N}{N}\int_x^1(1-t)^{(A+1)/2}(1+t)^{(B-3)/2}dt.
\end{multline}
Observe that
\begin{equation}
  \label{eq:167}
  \frac1{A+1} \binom{A+N}N = \frac1N \binom{A+N}{N-1},
\end{equation}
and the latter is an analytic function of all $A$.  Then both terms on
the right-hand side of~(\ref{eq:153}) are analytic functions of $A>-2$
for $-1<x\leq1$, so this last equation provides the analytic extension of
the integral~(\ref{eq:156}) defining $\delta\psi_N(x)$, which is \emph{sensu
  stricti} undefined for $A\leq-1$, to an analytic function on $A>-2$.

The rest of the reasoning is analogous to that in the orthogonal case.
The only technical difficulty arises because the error
term~(\ref{eq:err-hilb2}) in Szeg\H{o}'s formula does not depend on
$\theta$ in the range $0<\theta\leq c/N$, effectively making the reasoning of the
previous section inapplicable when $-2<A\leq-1$.  This is to be expected
since the summation formula only makes sense after being analytically
continued.  In what follows we prove that the various limits of the
kernel do in fact depend analytically on the parameter $A$, thus
allowing the expressions obtained for $A>-1$ to be extended to
$A>-2$.

Using Szego's formula~(\ref{eq:hilb2}) (valid for all $A$), there is
no problem to obtain this variant of lemma~\ref{lem:epsi}:
\begin{lemma} \label{lem:epsi2}
For any $A,B,\theta$ real and $0<\psi<\pi$ we have:
  \begin{equation}
    \label{eq:1000}
    \lim_{N\to\infty} N\int_{\theta/N}^\phi \psi_N^{(A,B)}(\cos\psi)\sin\psi\,d\psi = 2^{\frac{A+B}2}\int_\theta^\infty J_A.
  \end{equation}
\end{lemma}
\begin{lemma} \label{lem:epsi3}
  Using equation~(\ref{eq:153}), the expression
  \begin{equation}
    \label{eq:162}
    N\int_0^{\theta/N} \psi_N^{(A,B)}(\cos\phi)\sin\phi\,d\phi
  \end{equation}
  can be analytically continued to a regular function on $A>-2$.  As
  $N\to\infty$, this function tends to a limit which is also analytic for
  $A>-2$ and coincides with~(\ref{eq:168b}) for $A>-1$.
\end{lemma}
We change variables $\phi\to\phi/N$.  As before, we split the
integral to rewrite~(\ref{eq:162}) in the form
\begin{multline}
  \label{eq:163}
  2^{(A+B)/2}\int_0^\theta
  \left(\sin\frac\phi{2N}\right)^A\left(\cos\frac\phi{2N}\right)^B
  \left[\pAB_N\left(\cos\frac\phi N\right)-\pAB_N(1)\right]d\phi \\
  + 2^{(A+B)/2} \pAB_N(1)\int_0^\theta
  \left(\sin\frac\phi{2N}\right)^A\left(\cos\frac\phi{2N}\right)^B\pAB_N(1)\,d\phi
\end{multline}
The first of these terms is analytic for $A>-2$, the second one has an
analytic continuation given by~(\ref{eq:153}).  It is easy to see that
this second term has the asymptotic behavior:
\begin{multline}
  \label{eq:164}
  2^{(A+B)/2} \int_0^\theta
  \left(\sin\frac\phi{2N}\right)^A\left(\cos\frac\phi{2N}\right)^B\pAB_N(1)\,d\phi\\
  \sim 2^{\frac{B-A}2}\frac1N\binom{A+N}{N-1}\left(\frac\phi N\right)^{A+1}
\end{multline}
as $N\to\infty$, and from Stirling's formula~(\ref{eq:stir}), the binomial coefficient
$\binom{A+N}{N-1}=\frac{\G(A+N+1)}{\G(N)\G(A+2)}=O(N^{A+1})$, hence
this second terms is asymptotically negligible.  As for the first term
in~(\ref{eq:163}), we first write
\begin{multline}
  \label{eq:168}
  \pAB_N\left(\cos\frac\phi N\right) - \pAB_N(1)
  = -\frac1N \int_0^\phi{\pAB_N}'\left(\cos\frac\psi N\right) \sin\frac\psi N d\psi \\
  = -\frac{N+A+B+1}{2N} \int_0^\phi\pABp_{N-1}\left(\cos\frac\psi N\right) \sin\frac\psi N d\psi,
\end{multline}
where we have used the derivation formula~(\ref{eq:126}).  We can now
use Szeg\H{o}'s formula~(\ref{eq:hilb1}) to estimate $\pABp_{N-1}$
since $A+1>-1$.  The upshot is that the limit of~(\ref{eq:162}) as
$N\to\infty$ can be written as the following integral, which is an analytic
function of $A>-2$:
\begin{equation}
  \label{eq:1001}
  -2^{(A+B)/2}\int_0^\theta\int_0^\phi\phi^A\psi^{-A}J_{A+1}(\psi)d\psi\,d\phi.
\end{equation}
Using the Bessel function identity~(\ref{eq:121f}) we can simplify the
above integral, for $A>-1$:
\begin{equation}
  \label{eq:169}
  2^{(A+B)/2}\int_0^\theta J_A(\phi)d\phi,
\end{equation}
which is in agreement with lemma~\ref{lem:epsi}.

We note that the expression~(\ref{eq:169}) can be easily continued to
an analytic function of $A>-2$ without the need to rewrite it as the
double integral~(\ref{eq:1001}).  Namely, using~(\ref{eq:157}) we
have, for $A>-1$,
\begin{equation}
  \label{eq:170}
  \int_0^\theta J_A(\phi)d\phi = J_{A+1}(\theta) + \int_0^\theta J_{A+2}(\phi)d\phi.
\end{equation}
The expression on the right-hand side is analytic for $A>-2$ and
provides the desired analytic continuation.

The global level density is derived identically to the previous
section.  The limiting kernel in the bulk of the spectrum is given by
the sum of two terms: $\bar{S}^{(a)}_2(2\xi,2\eta)$ and another term
which is negligible in the limit.  For the central point $z=+1$, the
lemmas above yield the following expression for the limiting kernel:
\begin{equation}
  \label{eq:hS4}
  \hat{S}_4^{(a)}(\xi,\eta) = \sqrt\frac\xi\eta \hat{K}_2^{(A)}(2\xi,2\eta)
  - \frac\pi2 J_A(2\pi\eta)\int_0^{2\pi\xi}J_A(t)dt,
\end{equation}
whe+re the last integral is to be understood in the sense of
equation~(\ref{eq:170}) for $A\leq-1$.  Using equations~(\ref{eq:160})
and~(\ref{eq:152}) together with~(\ref{eq:161}) and the equation
above, the kernel can be rewritten in either of the forms:
\begin{align}
  \label{eq:163a}
  \hat{S}_4^{(a)}(\xi,\eta) &= \sqrt\frac\eta\xi \hat{K}_2^{(a)}(2\xi,2\eta)
  - \frac\pi2 J_{a-1}(2\pi\eta)\int_0^{2\pi\xi}J_{a+1}(t)dt,\\
  \label{eq:163b}
  \hat{S}_4^{(a)}(\xi,\eta) &= \sqrt\frac\eta\xi \hat{K}_2^{(a-2)}(2\xi,2\eta)
  - \frac\pi2 J_{a-1}(2\pi\eta)\int_0^{2\pi\xi}J_{a-3}(t)dt.
\end{align}

\bibliography{SymmRM}

\end{document}